\documentclass[12pt]{article}
\usepackage[utf8]{inputenc}
\usepackage{mathtools}
\usepackage[english]{babel}
\usepackage[T1]{fontenc}
\usepackage[font={small,it}]{caption}
\usepackage[a4paper,top=3cm,bottom=2cm,left=3cm,right=3cm,marginparwidth=1.75cm]{geometry}
\usepackage{adjustbox}
\usepackage{rotating}
\usepackage[font=footnotesize,textfont=it]{caption}
\usepackage{xcolor}
\usepackage{soul}

\usepackage{amsmath}
\usepackage{graphicx}
\usepackage[colorlinks=true, allcolors=blue]{hyperref}
\usepackage{wrapfig}
\usepackage{titlesec}

\titleformat{\section}
  {\normalfont\fontsize{14}{15}\bfseries}{\thesection}{1em}{}
  \titleformat{\subsection}
  {\normalfont\fontsize{12}{15}\bfseries}{\thesubsection}{1em}{}

\usepackage[backend=biber,style=apa
]{biblatex}

\title{How Cumulative is Technological Knowledge?}

\author{Peter Persoon\\
\texttt{P.G.J.Persoon@tue.nl}
  \and
   Rudi Bekkers\\
  \texttt{R.N.A.Bekkers@tue.nl}
  \and
  Floor Alkemade\\
  \texttt{F.Alkemade@tue.nl}
}
\date{November 2020}

\begin{document}

\maketitle
\begin{abstract}
Technological cumulativeness is considered one of the main mechanisms for technological progress, yet its exact meaning and dynamics often remain unclear. To develop a better understanding of this mechanism we approach a technology as a body of knowledge consisting of interlinked inventions. Technological cumulativeness can then be understood as the extent to which inventions build on other inventions within that same body of knowledge. The cumulativeness of a technology is therefore characterized by the \textit{structure} of its knowledge base, which is different from, but closely related to, the \textit{size} of its knowledge base. We analytically derive equations describing the relation between the cumulativeness and the size of the knowledge base. In addition, we empirically test our ideas for a number of selected technologies, using patent data. Our results suggest that cumulativeness increases proportionally with the size of the knowledge base, at a rate which varies considerably across technologies. At the same time we find that across technologies, this rate is inversely related to the rate of invention over time. This suggests that the cumulativeness increases relatively slow in rapidly growing technologies. In sum, the presented approach allows for an in depth, systematic analysis of cumulativeness variations across technologies and the knowledge dynamics underlying technology development.  
\end{abstract}

\section{Introduction}
\label{Introduction}

Technology progresses when engineers adapt their designs based on learning about previous designs. Consequently, a key element of theories of technological change is the cumulative nature of knowledge and invention: the idea that new results build on - or recombine - previous results \parencite{nelson_evolutionary_1982,freeman_economics_1997,basalla_evolution_1989,trajtenberg_university_1997}. Indeed, many of today's technologies have rich histories of development, some going back all the way to antiquity. While the size of the knowledge base of these technologies is substantial, this does not necessarily imply the underlying knowledge structure is cumulative: a pile of stones is different from a stone wall, and some walls are higher than others. 

Cumulativeness (or sometimes 'cumulativity') may therefore vary per technology and over time. A better understanding of the underlying mechanisms of technological cumulativeness is important for a number of reasons. From an economics perspective, the extent to which a technology develops in a cumulative manner has implications for how easy it is to enter or diversify into that technology. Entry is considered more difficult in complex technologies that require extensive and in-depth knowledge about the underlying principles \parencite{winter_schumpeterian_1984, breschi_technological_2000,breschi_geography_2000}. Recent contributions from the geography of innovation describe how regions are more likely to diversify into technologies that are related to their existing knowledge base \parencite{balland_relatedness_2016, boschma_relatedness_2015,balland_geography_2017}. An understanding of the cumulative nature of technological development is thus pivotal for ongoing efforts of smart specialization \parencite{foray_smart_2014}, where regions seek out attractive technologies for future specialization. From a philosophical perspective, a better understanding of cumulativeness and its role in the evolution of technological knowledge \parencite{arthur_nature_2009} may help to clarify the relation between knowledge accumulation and the complexity of that knowledge, which is an ongoing discussion in the 'cumulative culture' literature \parencite{vaesen_complexity_2017,tennie_ratcheting_2009, dean_human_2014}. Developing this understanding starts from a clear definition and measure of cumulativity.

Surprisingly, despite the recognized importance of cumulativity, the exact meaning of the concept often remains unclear. Characterizations vary from the incremental change in artifacts  \parencite{ogburn_social_1922, gilfillan_sociology_1935, butler_notebooks_2014, basalla_evolution_1989}, to the persistence of innovative activity \parencite{malerba_technological_1993, cefis_is_2003, suarez_persistence_2014}, to the building of technological knowledge on earlier findings \parencite{trajtenberg_university_1997, merges_limiting_1994, scotchmer_standing_1991,enquist_modelling_2011}. 

In this contribution we aim to develop a better understanding of technological cumulativeness by taking the following steps: In Section \ref{Theory} we present a comprehensive review of the various perspectives on cumulativeness and identify their common grounds. In Section \ref{Indicators} we use this analysis to formulate two indicators which measure cumulativeness: the \textit{internal dependence} and \textit{internal path length}. In Section \ref{dynamics} we then discuss how the values of these indicators are expected to change as a technology develops. In Section \ref{EmpiricalAnalysis} we test these expectations empirically for a number of technologies, using patent data as a proxy for inventions. Finally we discuss some deeper implications of our contribution to the understanding of technological cumulativeness in Section \ref{Discussion} and summarize our main conclusions in Section~\ref{conclusions}. 

\section{Theoretical perspectives on technological cumulativeness} 
\label{Theory}

Where in most texts 'cumulative' simply means 'summed up', in the innovation literature the term has come to represent a type of technological development. Perspectives on cumulative technological development however vary across contributions.

The earliest ideas about technological cumulativeness arise in studies of the gradual change in pre-20th century artifacts \parencite{pitt-rivers_evolution_2018,butler_notebooks_2014,gilfillan_inventing_1935}, which are reminiscent of fossil records of gradually evolving species. Inspired by evolutionary theory, these theories understand technological change as a process in which antecedent artifacts are \textit{replicated with incremental modifications}, thereby creating descendant artifacts \parencite{gilfillan_sociology_1935,ogburn_social_1922}. In this first perspective, artifacts are literally the sum of many incremental modifications, justifying the term 'cumulative'. 

While the cumulative aspect of technology arises naturally in this perspective, it is unclear when a development is \text{not} cumulative: as in genetic lineage, \textit{each} descendent is supposed to have an antecedent. Some authors have argued that in reality, technological developments occasionally 'jump'; when a radical finding breaks fundamentally with past engineering practices and ideas~\parencite{schoenmakers_technological_2010, verhoeven_measuring_2016} it may initiate a new model of solutions to selected technological problems, i.e., a new \textit{technological paradigm} \parencite{dosi_technological_1982}. In this second perspective, cumulative development is \textit{the opposite of radical development}, and interpreted as the incremental change happening within a technological paradigm.  

Yet, to base cumulative change solely on the notion of incremental change raises two difficulties. First, there is a certain arbitrariness to when a change is incremental or not. Depending on context and knowledge of the subject, different people may characterize incrementality differently. Second, even if the change from an antecedent to descendant is radical, the antecedent may still be of crucial importance to the formation of the descendant \parencite{basalla_evolution_1989}. 

These difficulties are sidestepped in a third perspective, where a development is cumulative if a later result \textit{depends} or \textit{builds on} an earlier result~ \parencite{merges_limiting_1994, breschi_technological_2000, trajtenberg_university_1997,enquist_modelling_2011}. 'Dependence' or 'dependency' is here interpreted in the context of technology as a body of knowledge, where new technological ideas or inventions (the 'results') draw on earlier insights, and are themselves used in later ideas and inventions. Note in this perspective, cumulativeness is a property of the \textit{development} (not of one of the results). If we are  interested in the cumulativeness of a technology, we therefore consider all developments within that technology, i.e. all dependencies between results that are part of that technology. Alternatively, authors have studied the cumulativeness of the union of multiple (or all) technologies \parencite{acemoglu_innovation_2016,napolitano_technology_2018,clancy_combinations_2018}, thereby focusing on inter-technology developments or dependencies. Both approaches are relevant to better understand the advancement of technology and knowledge production. In this work we however focus on the former approach, as we are mainly interested in question to what extent cumulativeness is an intrinsic property of a technology, and how this property varies for different technologies.  

The relevance of cumulativeness as an intrinsic property of a technology is reflected by its role as defining element of a \textit{technological regime} \parencite{nelson_evolutionary_1982}, which defines the relevant circumstances under which innovating firms or organisations compete, thrive or fail. Within a technological regime, higher cumulativeness is associated with greater appropriability of innovation and greater (geographical) concentration of innovative activity \parencite{winter_schumpeterian_1984,malerba_schumpeterian_1996, breschi_technological_2000}. The framework of technological regimes gave rise to a number of contributions which use yet another perspective of cumulativeness, where the emphasis is not so much on the dependence of later generations of a technology on earlier ones, but more on the continuation of those generations \parencite{malerba_persistence_1997,cefis_is_2003,frenz_what_2012, apa_knowledge_2018,holzl_distance_2014, breschi_geography_2000}. Cumulativeness is then characterized by the \textit{persistence} of inventive and innovative activity in a technology: the longer a development continues (without significant interruption), the greater the cumulativeness. Where previous perspectives focus more on cumulativeness as an intrinsic property of technology, this fourth perspective also attributes a role to the creators of the technology (and their persistence to continue along a given path). 

In summary, we recite from these four different perspectives the key notions of technological cumulativeness: (1) as replication with incremental modifications, (2) as within-paradigm (opposite to radical) development, (3) as dependence or building on earlier technology and (4) as persistence of inventive or innovative activity. The first two perspectives approach cumulativeness as 'incremental change', the latter two perspectives approach cumulativeness as 'continuous dependence' of technology on earlier generations of technology. Though apparently very different, there are similarities between incremental change and continuous dependence. Incremental change supposes a series of modifications to what is, in some sense, a single object (often pictured as an artifact). Similarly, continuous dependence supposes a series of dependencies between objects which are, in some sense, different (often pictured as set of inventions). Essentially therefore, the discrepancy is about the object(s) to which a series of changes is applied, yet both advocate the relevance of \textit{a series of developmental steps}. Further, both for incremental change and continuous dependence, cumulativeness appears in two dimensions: (i) the size of each developmental step: if the modification is small (dependence is great), the cumulativity is large and (ii) the number of steps in the process: if there are many small modifications (a long chain of dependency links) the cumulativity is large. While (i) and (ii) both relate to cumulativity, they are theoretically very different, and we shall henceforth refer to them as the \textit{transversal-} and \textit{longitudinal dimension} of cumulativity respectively. Although both dimensions can be meaningfully interpreted in all four cumulativeness perspectives, it appears the first two perspectives focus more on the transversal dimension and the latter two perspectives more on the longitudinal dimension. In the next section we will propose a separate indicator for each dimension. We emphasize that both are measured \textit{within} a certain technological field or technology. Although the interaction between multiple fields or technologies is interesting and worth studying, the focus of this work is on understanding these cumulativeness dimensions within a single technology. 

Finally we discuss the relation between technological cumulativeness and complexity. In this contribution we will not enter the discussion about the exact meaning of technological complexity (for a good overview see \cite{vaesen_complexity_2017}), but instead work with the general description of a complex system consisting of many, non-trivially interacting subsystems \parencite{simon_architecture_1962}. One way to interpret this in the context of technology, is to consider an invention to be a system consisting of subsystems, which are (parts of) other inventions or borrowed ideas. The complex character of an invention is therefore in an abstract sense captured by the transversal dimension of cumulativeness, which focuses on these direct dependencies. Intuitively, the more subsystems and dependencies, the greater the complexity (although this strongly depends the chosen measure for complexity). However, this is not the entire story. A relevant criterion for increasing complexity in the context of evolutionary systems is that a representative sample of lineages of descent increases in complexity \parencite{mcshea_complexity_1991,vaesen_complexity_2017}. Not only therefore should 'more complex' systems appear in time, but these should also fit into the lines connecting antecedents and descendants. In the context of technological knowledge, the lines of descent appear rather literally in the mentioned first perspective of cumulativeness, and correspond to the longitudinal dimension of cumulativeness. Especially the joint consideration of the transversal and longitudinal dimensions of cumulativeness therefore allows us to study the dynamics of technological complexity. 

\section{Measuring cumulativeness}\label{Indicators}

In most contributions mentioning cumulative technological development, cumulativeness remains an abstract property without explicit measure. There are a number of exceptions however, in particular the contributions adhering to the earlier mentioned 'persistence perspective' of cumulativeness. These contributions base their measures of cumulativeness on a variety of sources: survey data  \parencite{breschi_technological_2000, frenz_what_2012, holzl_distance_2014}, licensing data \parencite{lee_effects_2017} and  statistical properties of patent count time series \parencite{malerba_persistence_1997, cefis_is_2003, breschi_geography_2000}. While all of these highlight interesting aspects of cumulative processes, none of them seem to directly proxy the key property of knowledge building on knowledge. Survey data may offer detailed information on usage of particular knowledge, yet it is challenging to quantify and generalize this information in order to compare different technologies. Approaches based on counting backward citations \parencite{apa_knowledge_2018} arguably do measure the extent to which knowledge builds on earlier knowledge, yet without specifying \textit{which} technologies are cited, only partially capture the underlying knowledge structure of technologies. However, as was argued in the previous section, to understand technological cumulativeness along both the transversal and longitudinal dimension, studying the underlying knowledge structure is pivotal. In this contribution our starting point is to interpret this structure as a network of interconnected elements of knowledge. Each node then represents a single invention, and each link represents a knowledge flow. A link thus naturally corresponds to a dependence, or knowledge building on other knowledge. This approach has been successfully applied to the analysis of breakthrough innovation \parencite{dahlin_when_2005, verhoeven_measuring_2016, fleming_recombinant_2001}, main paths \parencite{hummon_connectivity_1989,verspagen_mapping_2007}, emerging technologies \parencite{shibata_comparative_2009,erdi_prediction_2013} and technological network evolution \parencite{valverde_topology_2007}. We denote the knowledge flows \textit{to} an invention (i.e., the links which indicate on which knowledge the invention builds) as 'backward links' and the knowledge flows \textit{from} an invention as 'forward links'. 

Further, we assume that there is a technology classification which allows us to assign each invention to at least one class, hence allowing us to distinguish between {\it internal links} (link to an invention in the same class) and {\it external links} (link to an invention of another class)\footnote{Inevitably, there is some room for interpretation here as there can be various grounds on which technologies are classified. In the Section \ref{Discussion} we discuss a number of alternative approaches to making the external-internal distinction.}. In the previous section we introduced the transversal and longitudinal dimensions of cumulativeness. 
Exploiting useful network structures, we will in the next two subsections introduce two indicators measuring the cumulativeness along these dimensions. For the transversal dimensions we introduce the internal dependence and for the longitudinal dimension we introduce the internal path length. 

\subsection{The transversal dimension: Internal dependence}

The transversal dimension of cumulativeness reflects the extent to which findings in a given technology \textit{depend} on other findings within that technology. In a network of inventions, each directed link can rather literally be interpreted as a relation of dependence. Ideally, we would go into the content of each knowledge link to distinguish a degree of dependence. Yet this approach would be difficult to automate when the number of links and inventions becomes large (which is the case for most technologies). Most network approaches to technology therefore count each knowledge link equal, so the number of internal links becomes a measure for the dependence. Each invention which is added to the technology introduces a number of backward internal links, see Figure \ref{id_and_ipl} (left panel) for a network illustration. The more internal backward links it introduces, the more the technology builds on itself. As a measure for the transversal dimension, we therefore define the \textit{internal dependence (id)} of a technology as the \textit{average number of backward internal links per invention}. A high id signals high cumulativity in the transversal dimension. 
\subsection{The longitudinal dimension: Internal path length}
\begin{wrapfigure}{r}{0 pt} 
\centering
\includegraphics[width=0.5\textwidth]{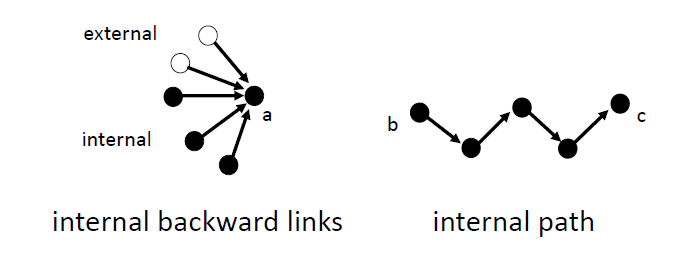}
\caption{\textbf{Useful network structures} Left: The number of internal backward links of node \textit{a} is 3. Right: the length of the internal path between node \textit{b} and {c} is 4. For a precise definition of path and path length we refer to section \ref{ipl_section}.}
\label{id_and_ipl}
\end{wrapfigure}

The longitudinal dimension of cumulativity reflects the number of steps in a series of technological developments. Approaching technology as a network of inventions, we can translate this rather literally to a chain of internal inventions connected by links, which translates to the notion of a 'path' in the terminology of network analysis, see Figure \ref{id_and_ipl} (right panel) for a network illustration. The longer the internal paths, the longer a series of developments within a technology is continued. As multiple knowledge aspects of a technology may develop in parallel, we generally deal with several, intertwined paths. As a measure for the longitudinal  dimension, we therefore define the \textit{internal path length (ipl)} of a technology as the \textit{average length of all paths within that technology}. A high ipl signals high cumulativity in the longitudinal dimension.\footnote{Similar ideas are presented in \parencite{frenken_branching_2012}, where innovations attain 'higher quality' with longer path lengths. The study of Frenken et al., which is based on numerical simulations, thereby focuses on the (re)combination principle in relation to diffusion.}

\section{Modeling the knowledge dynamics}\label{dynamics}  
In this section we discuss how the values of the internal dependence (id) and internal path length (ipl) are expected to change as a technology develops, i.e. when the size of its knowledge base increases. More specifically, we analyze  (a) how the id and ipl change as the number inventions increase and (b) how the id and ipl are interrelated. We thereby describe both general and technology-specific elements.

\subsection{Invention as search process}

In this section we sketch a highly simplified model of the invention process in a certain technology, which consists of an inventor performing a series of searches. Essentially, the inventor searches until he or she succeeds in completing an invention, where a knowledge flow (equivalent to a backward internal link) is picked up along with each search. The relevant quantity in this process is the probability $\rho$ of completing an invention before performing another search, which may depend on the size of the knowledge base of a technology, as measured by the number of inventions $n$. For each $n$, the probability of inventing is therefore $\rho(n)$, the probability for performing a search is $1-\rho(n)$. We have two main assumptions in this model: 
\begin{enumerate}
    \item The probability $\rho(n)$ decreases proportionally with the number of inventions $n$. This reflects the intuition that it becomes harder, as a technology develops, to produce an invention without using any prior knowledge developed in that technology. In other words, the inventor needs to consider some knowledge in a certain field before delivering a contribution to that field, and the larger the field, the more the inventor needs to consider. 
\item The probability of success is independent of the number of searches: in the invention process, there is no guarantee that a certain amount of effort leads to success.
\end{enumerate} 
Given these assumptions we may write down for the probability $\rho(n)=\frac{1}{qn+m_{1}}$, introducing the technology specific constants $q>0$ and $m_{1}>1$. Here, the parameter $m_{1}$ describes the need to have knowledge of the technology in order to invent at the initial stage of this technology, and $q$ describes how fast this need increases as the technology develops. As a consequence of the two assumptions, the probability for a node to have $m$ backward internal links (i.e. the probability that $m$ searches take place before invention) is given by $P_{n}(m)=(1-\rho(n))^{m}\rho(n)$, i.e. the number of backward internal links per node is distributed geometrically.\footnote{We then assume that the number of backward links per node stays well below $n$, which appears to be reasonable if we consider technologies with large $n$.} This distribution is characterized by a highly skewed shape towards lower values of $m$, yet as $n$ increases, it slowly becomes less skewed. 

The rate $q$ is related to the type of technological knowledge and we therefore assume it is a technology specific quantity. Yet we hypothesize it is also related to the rate of invention over time. Our reasoning is as follows. If the rate of invention over time is high, this means that more people work on the same technology at the same time. If multiple researchers work on the same technology, they tend to specialize, focusing only on a particular sub-field or sub-part of the technology. As an effect, multiple aspects of the technology develop in parallel, perhaps more so than if a smaller group of people had worked on it. As a result the development of the technology is more fragmented into sub-fields, which causes inventors active in these sub-fields to focus on the relevant findings within their sub-field. We may therefore suppose that there is structurally less need for these inventors to master the entire knowledge base, which leads to lower values of $q$. In reverse, it is possible that a low need for prior knowledge of a technology accelerates innovative activities in a technology, as it may then be more easily accessible, thus inviting more people to contribute. Deriving a more precise form and causal direction of the inverse relation between $q$ and rate of inventing over time however is beyond the scope of this work. For a more elaborate discussion of the causality we refer to Section \ref{Discussion}.

\subsection{Internal Dependence Dynamics}\label{id_section}

Using the distribution of the number of backward internal links, we can calculate $\langle m \rangle=\sum_{m=0}^{n}mP_{n}(m)$ the expected value of the number of backward internal links per invention, i.e. the internal dependence (id). Assuming that $n$ is large, we can approximate this sum by choosing infinity for the upper limit and using the expression $P_{n}(m)=(1-\rho(n))^{m}\rho(n)$, obtaining 
\begin{equation}\label{id_linear}
    \langle m \rangle=\frac{1}{\rho(n)}-1=qn+m_{1}-1=qn+m_{0},
\end{equation}
introducing $m_{0}=m_{1}-1$ for convenience. We therefore conclude that the id is expected to increase proportionally with the number of inventions (i.e. with the size of the knowledge base), where the rate can by approximated by $q$ for a large number of inventions. This technology specific coefficient $q$ describes how fast the need to have specialized knowledge increases in order to produce an invention in that technology. 

\subsection{Internal Path Length Dynamics}\label{ipl_section}

Next we will discuss how we expect the internal path length (ipl) to depend on the number of inventions. Although these results can be generalized by including external links, we focus in this contribution for simplicity on the role of internal links. A new invention creates at least one new path with each of its internal backward links. The internal dependence, besides measuring a complementary dimension of cumulativity, therefore also plays a key role in the ipl dynamics. Let us again consider a technology with $n$ inventions, where the $n$th invention has \textit{on average} $\langle m \rangle$ internal backward links. Some inventions however will have no backward links, which we will refer to as \textit{initial inventions}. As a first assumption, we take that the number of initial inventions $n_{0}$ is a fixed fraction $r$ of $n$, i.e., $n_{0}=rn$.\footnote{As we explain in more detail in Appendix \ref{math} this assumption is compatible with the found backward link distribution if $q$ is small compared to $m_{0}$, we then have that $r\approx 1/(m_{0}+1)$.} We use the initial inventions to define a path and path length: \begin{itemize}
    \item[-] A \textit{path} is a sequence of inventions $\textbf{i}_{0},\textbf{i}_{1},..., \textbf{i}_{k}$ in which for any $k\geq 0$ and $x>0$, $\textbf{i}_{x}$ has a backward link to invention $\textbf{i}_{x-1}$ and $\textbf{i}_{0}$ is an initial invention. 
\item[-] The \textit{path length} of path $\textbf{i}_{0},\textbf{i}_{1},..., \textbf{i}_{k}$ is $k$. 
\end{itemize}
We denote the number of paths of length $k$ by $f_{k}(n)$. From the first assumption\footnote{If we also consider external inventions, we can choose a more general definition, where a path can also start at an external invention. Note that, ignoring the links to external inventions, the inventions which only link to external inventions become initial inventions}, we have that, $f_{0}(n)=rn$. As a second assumption, each invention is equally likely to be used as prior knowledge with probability $\frac{1}{n}$. Let us consider what happens to $f_{k}(n)$ for $k>0$ when we introduce the $n+1$th invention. If that invention builds on a prior invention $\textbf{i}$ that has $l_{i,k-1}$ paths of length $k-1$, each of these paths will increase by 1, hence $f_{k}(n)$ increases by $l_{i,k-1}$. This holds for all inventions, which in total have $\sum_{i}l_{i,k-1}=f_{k-1}(n)$ paths of length $k-1$. For $\Delta_{n} f_{k}(n)$, i.e. the expected increase in $f_{k}(n)$ from $n$ to $n+1$, we therefore have $\Delta_{n} f_{k}(n)\propto f_{k-1}(n)$, and for $k>1$ we have
\begin{equation}\label{main1}
    \Delta_{n} f_{k}(n)=\langle m \rangle\frac{f_{k-1}(n)}{n}.
\end{equation}
In the previous section we established that $\langle m \rangle \approx qn+m_{0}$. When $n$ gets large, $\langle m \rangle/n\to q$, further reducing Equation \ref{main1} to 
\begin{equation}\label{main3}
    \Delta_{n} f_{k}(n)=q f_{k-1}(n).
\end{equation}
As there are no paths for $n=0$, we take that $f_{k}(0)=0$ for all $k$. Using this initial condition and the expression for $f_{0}(n)$, the solution to Equation \ref{main3} is derived to be
\begin{equation}\label{main2}
    f_{k}(n)=rq^{k}\binom{n}{k+1}, 
\end{equation}
where $\binom{x}{y}$ is the binomial coefficient. The steps leading to this solution and later ones are explained in more mathematical detail in Appendix \ref{math}. Summing over all $k$ we obtain the total number of paths $\sum_{k=0}^{n}f_{k}(n)=r(1+q)^n/q-r/q$. The total number of paths is therefore expected to increase exponentially in $n$. For the normalized path length distribution $\tilde{f}_{k}(n)$, describing the probability to have a path of length $k$, we subsequently obtain 
\begin{equation}\label{main4}
    \tilde{f}_{k}(n)=\binom{n}{k+1}\frac{q^{k+1}}{(1+q)^n-1}, 
\end{equation} 
which is a distribution closely related to the binomial distribution. This indicates that as $n$ increases, the path length distribution will shift from a skewed shape towards more symmetric, parabolic shape (on a log scale) and its maximum, the most frequent path length, will continuously shift to higher values. Subsequently, we can calculate the expected path length $\langle k \rangle=\sum_{k=0}^{n}k\tilde{f}_{k}(n)$, i.e. the ipl, which reduces for large $n$ to 
\begin{equation}\label{slope}
    \langle k \rangle\approx \frac{q}{q+1}n+k_{0},
\end{equation}
where $k_{0}$ is some constant value. As we focus on large $n$ behavior, we are less interested in this constant. What is more important is the expectation that the ipl increases proportionally with the number of inventions, by a rate $p=q/(q+1)$. This rate $p$ is a number between $0$ and $1$: for large $q$ it is close to 1 and for small $q$, it is close to $q$. We end this section by mentioning two extensions of the model which improve its explanatory power. 
\begin{itemize}
    \item In this derivation, we assumed that $\langle m \rangle/n\approx q$, even though we know it in fact only \textit{approaches} $q$ for large $n$. This approximation can be significantly improved by instead calculating the average $\langle m \rangle/n$ for $n$ inventions. We can determine this quantity in two ways, (a) by directly using the data of the number of backward links for each invention, i.e. by calculating  $q'_a=1/n\sum_{i}^{n}m_{i}/i$ where $m_{i}$ is the number of backward links of invention $i$ and (b) by using estimates for parameters in the relation $\langle m \rangle=m_{0}+nq$, i.e. calculating $q'_{b}=1/n\sum_{i}^{n}q+m_{0}/i=q+m_{0}H(n)/n$, where $H(n)$ is the $n$th harmonic number. Analogous to Equation \ref{slope}, we then have $q'_{a}/(1+q'_{a})=p'_{a}$ and similar for $p'_{b}$. $p'_{a}$ Is likely to be more accurate as it is more directly based on the backward link data, yet $p_{b}'$ is less sensitive to outliers in this data. Both predictions should however be close to one another. Note that this correction depends proportionally on $m_{0}$. 
    \item Equation \ref{main1} implies that as we add the $n^{th}$ invention to the system, the number of paths of length $n$ increase from $0$ to some positive value. In fact this equation therefore establishes a 'maximum speed' $v$ of $1$ path length per invention, faster than which the path lengths cannot increase. This maximum speed is rather lenient: technologies with paths increasing with 1 length per invention (i.e. forming perfect chains) would be highly unrealistic. While Equation \ref{main1} is accurate for the more frequent path lengths (i.e. the lengths close to the mean), it may therefore be less accurate for the less frequent path lengths (i.e. the shortest and longest lengths). A more realistic estimate of the maximum speed $v$ may therefore help establish a better description of the overall distribution of path lengths. Let us suppose that we at once add $\delta n$ inventions to the system which do not connect amongst themselves, and of which the total added number of backward links is $M(n)$. Equation \ref{main1} then becomes
    \begin{equation}\label{main5}
        f_{k+1}(n+\delta n) - f_{k+1}(n)=M(n)\frac{f_{k}(n)}{n}
    \end{equation}
    If we choose $\delta n$ such that $M(n)\approx n$, then each of the $n$ inventions in the system approximately obtains $1$ forward link. This implies that all paths in the system increase on average by $1$, including the longest path(s). $\delta n$ Therefore defines a typical interval for the longest path to increase by $1$, and $1/\delta n$ therefore presents a more reasonable estimate for the maximum speed $v$. We will use this idea to derive a new expression for the path length distribution. Note that Equation \ref{main5} then becomes
    \begin{equation}\label{main6b}
        f_{k+1}(n+\delta n)-f_{k+1}(n)=f_{k}(n).
     \end{equation}
   If we introduce the variable $n'=n/\delta n$ and the function $f'_{k}(n')=f_{k}(n)$, we may write this relation as 
 $f'_{k+1}(n'+1)-f'_{k+1}(n')=f'_{k}(n')$, which is solved by $f'_{k}(n')=r\binom{n'}{k+1}$ (this time using the condition that $f'_{k}(n')=0$ for $k<n'=nv$). This leads to the normalized distribution 
\begin{equation}\label{main7}
\tilde{f}'_{k}(n')=\frac{1}{2^{n'}-1}\binom{n'}{k+1}
\end{equation}
and expected path length (i.e. the ipl)
\begin{equation}\label{main6}
   \langle k \rangle' \approx \frac{n'}{2}+k_{0}',
    \end{equation}
       where $k_{0}'$ is again a constant we are less interested in. Rewriting this expression in terms of $n$ gives the coefficient $\frac{1}{2\delta n}$ or $\frac{v}{2}$, describing how fast the ipl increases with $n$. Assuming the earlier analysis with a greater maximum speed is accurate for the mean path length values, this should coincide with the earlier established coefficient $p$. We can therefore approximate the maximum speed as $v\approx 2p$.\footnote{This is consistent with the earlier assertion that $M(n)\approx n$. To see this, note that the total number of links is $n\langle m \rangle$ (as $\langle m \rangle$ is an average), hence between $n$ and $n+\delta n$ we add $\delta n(m_{0}+q(\delta n+2n))$ links. For this to equal $n$ in the limit where $n$ becomes large, we require $\delta n\rightarrow \frac{1}{2q}$. In the same limit, $p\rightarrow q/(q+1)$, which is approximately $q$ for small $q$. This is therefore consistent with $\frac{1}{\delta n}=v \approx 2p$}
This implies that the paths with maximum length grow about twice as fast as paths with mean length, i.e. the distribution becomes more symmetric as $n$ increases. Noting that $n'=nv$, we identify $n'$ as the maximum path length after $n$ inventions, which can be used to evaluate Equation \ref{main7}. Alternatively, we use the expression for $v=2p$ to rewrite this expression in terms of $n$ and $p$,
\begin{equation}\label{main8}
\tilde{f}_{k}(n)=\frac{1}{4^{pn}-1}\binom{2pn}{k+1}.
\end{equation}
\end{itemize}
\section{Empirical analysis}
\label{EmpiricalAnalysis}

In this section we empirically test the models developed in Section \ref{dynamics} using patent and patent citation data. First, we discuss our type of data and a number of limitations of these data. Subsequently, we perform the analysis on three different levels: first we consider the development and distributions of both cumulativeness indicators for four focus technologies into detail. Second, we consider the relation between the two indicators and the consistency of the indicators, using a larger set of technologies. Third, we choose a more aggregated level of technology classification to obtain a more general overview of the cumulativeness variation across different technological fields, which also allows us to compare our findings to earlier results from the literature and to some extent validate the indicators.   

\subsection{Data description}

In order to study the knowledge dynamics empirically, we need some codification of that knowledge. Patents are an important codification of technological knowledge, as each patent is a detailed description of a new, non-trivial technological development. Furthermore, patent systems have two elements that allow us to study technological content without necessarily having to consider the detailed meaning of each individual patent. The first element is that of patent citations, which identify one to one, directional content relations between patents. This enables us to study the flow of knowledge \parencite{jaffe_characterizing_1989,jaffe_geographic_1993}. The second element is that of the patent classifications, which hierarchically groups patents on the basis of their content. This enables us to focus specifically on the development of a particular technology, distinguishing between internal and external knowledge. A basic assumption of our work is that cumulativeness is an intrinsic property of technology, which is independent from the way the technology is patented. It is therefore important to keep in mind the limitations of representing technological knowledge by patent data, which will henceforth discuss. For each limitation we mention how we attempt to account for it. 
\begin{enumerate}
    \item Not all technology is or can be patented, \parencite{jaffe_adam_b._patent_2017} and the 'quality' of patents (evaluated against the patentability requirements) varies \parencite{de_rassenfosse_low-quality_2016, jaffe_innovation_2004}. Especially when the number of patents involved is small, without a detailed examination of the content we risk misrepresenting a technology. In this analysis we therefore choose technologies for which the number of patents is relatively large. Also, we only consider \textit{granted} patents, which have withstood the critical assessment of patent examiners. 
    \item Citations may not always represent actual knowledge flows \parencite{criscuolo_does_2008}. Citations may be provided by inventors but may also be added by examiners, and while the first may be more indicative for knowledge flow, the distinction was not always documented by all patent offices \parencite{azagra-caro_examiner_2018}. We therefore include an additional analysis in Appendix \ref{appendix_5} of the effect of both types of citations (examiner or inventor added) to the knowledge dynamics.
    \item There are institutional differences between patent offices around the globe, which may affect the way inventions and linkages to prior art are documented. \parencite{bacchiocchi_international_2010}. An important difference is for example is the greater tendency to cite in the United States patent system than in the European patent system \parencite{criscuolo_does_2008}, which may impact the value of our indicators. To account for these differences we therefore do this analysis for patents from two different patent systems, choosing the US system (organized by the US patent office USPTO) and European  system (organized by the European Patent Office EPO).  
\end{enumerate}
To aggregate patents of which the technological content is the same, we choose a patent family as a basic unit or node, creating a US data set selecting families with at least one USPTO member and a European data set selecting families with at least one EPO member\footnote{To be precise, we choose the DOCDB type of patent family, where all family members have exactly the same priorities}. In the US data set each unique reference (backward citation) of a US member of each family to any member of another family in our data set represents a unique link (hence we do not limit our selection to US-US citations only).\footnote{Note that if we had selected \textit{any} family citation we effectively take the union of all citations, hence failing to distinguish between the citing tendencies of different patent systems.} Our European data set is created analogously. Henceforth by 'US patent' we actually refer to an patent family containing a US member which is granted, and similar for 'European patent' or 'EP patent'. 

In order to select and demarcate technologies, we used the Cooperative Patent Classification (CPC) \parencite{cpc_cooperative_2018}. In this analysis we consider technologies on two levels of classification: the CPC group/subgroup level and a more aggregated level of classification. For the group/subgroup analysis we choose a set of 24 arbitrary technologies, yet making sure that (i) the set is diverse (including technologies from each main CPC section and from mostly different subclasses) and (ii) each technology contains a reasonably large number of patents (for US $>$700 and EP $>$200). Table \ref{desstat} and Table \ref{addtech} indicate the CPC codes and number of patents of these technologies. Table \ref{desstat} singles out four 'focus technologies' which we will analyze in more detail. The sub-selection of the focus technologies was made choosing considerable variation in (a) knowledge base size (where nuclear fission has 3608 US patents, photovoltaics has over 9000), (b) age (where nuclear fission started developing in the 1960's, the main development of wind turbines starts from the 1990's), (c) the working (theoretical) principles behind the technologies (varying from nuclear physics to aerodynamics). From both Table \ref{desstat} and \ref{addtech} it is clear there are generally more US than European patents, even taking into account that the EP patents do not go back further than 1978. As the column with the number of patents in the same family indicates, most European patents (around 75 percent) have a US equivalent as well. 

For the more aggregated level of classification we grouped together patent classes analogous to the approach by Malerba and Orsenigo \parencite{malerba_schumpeterian_1996}. However, given that their publication now dates more than 20 years back, and the patent classification system is subject to constant change, some differences between their grouping of classes and ours is inevitable\footnote{As a matter of fact, the CPC  did not yet exist at the time of the Malerba and Orsenigo paper, yet the closely related International Patent Classification (IPC) did.}. In Table \ref{super_class} in Appendix \ref{appendix_3} we present an overview of our grouping, note that we take the union of CPC classes (hence counting each patent once). The data in this research comes from the Patstat 2019 spring edition. 
\begin{table}[h!]
\centering
\caption{
\textbf{Description of the four focus technologies.} The selected patents have an earliest filing year<2009.}
\scriptsize
\resizebox{\textwidth}{!}{%
\begin{tabular}{l|l|l|l|l|l}
\textbf{\begin{tabular}[c]{@{}l@{}}Technology \\ short name\end{tabular}} & \textbf{CPC code} & \textbf{CPC description} & \textbf{\begin{tabular}[c]{@{}l@{}}\#US \\ granted \\ patents\end{tabular}} & \textbf{\begin{tabular}[c]{@{}l@{}}\# EP \\ granted\\ patents\end{tabular}} & \textbf{\begin{tabular}[c]{@{}l@{}}\# same\\ family \end{tabular}} \\ \hline
Nuclear Fission & Y02E 30/3 & \begin{tabular}[c]{@{}l@{}}Energy generation of nuclear \\ origin: nuclear fission reactors\end{tabular} & 3608 & 745
 & 558 \\ \hline
Photovoltaics & Y02E 10/5 & \begin{tabular}[c]{@{}l@{}}Energy generation through\\  renewable energy sources: photovoltaic energy\end{tabular} & 9088 & 2599 & 1947 \\ \hline
Wind Turbines & Y02E 10/7 & \begin{tabular}[c]{@{}l@{}}Energy generation through\\  renewable energy sources: wind energy\end{tabular} & 5405
 & 1767 & 1323\\ \hline
Combustion Engines & F02B 3/06 & \begin{tabular}[c]{@{}l@{}}Engines characterised by air \\ compression and subsequent fuel \\ addition with compression ignition\end{tabular} & 6466
 & 2089 & 1344 \\ \hline
\end{tabular}
}
\label{desstat}
\end{table}
Time is not adopted as an explicit variable in our models, yet we check for the consistency of our models over time and at a later point consider the invention rate over time. We do that by using the earliest filing date of the patent, as it is the closest point in time to the actual invention and therefore helps to establish a chronological ordering of inventions. It generally takes several years however before filed patents are actually granted: the European patents granted in 2012 were on average first filed 6.5 years earlier, for US patents this was about 5 years. Likewise, from all patents eventually granted which were filed earliest in 2005, it took 50 percent 6.9 years to be granted, and it took about 12.5 years for 95 percent of them to be granted. For US patents filed earliest in 2005 the same percentages correspond to about 5 and 10 years respectively. To be relatively confident to include 95 percent of the patents for each year considered, hence avoiding a 'truncation effect' as much as possible, calculating back from 2019, we should therefore not consider earliest filing years later than 2008. 

\subsection{Id and ipl for the focus technologies} 

\begin{wrapfigure}{ht}{0.6\textwidth} 
\centering
\includegraphics[width=0.6\textwidth, height=0.5\textwidth]{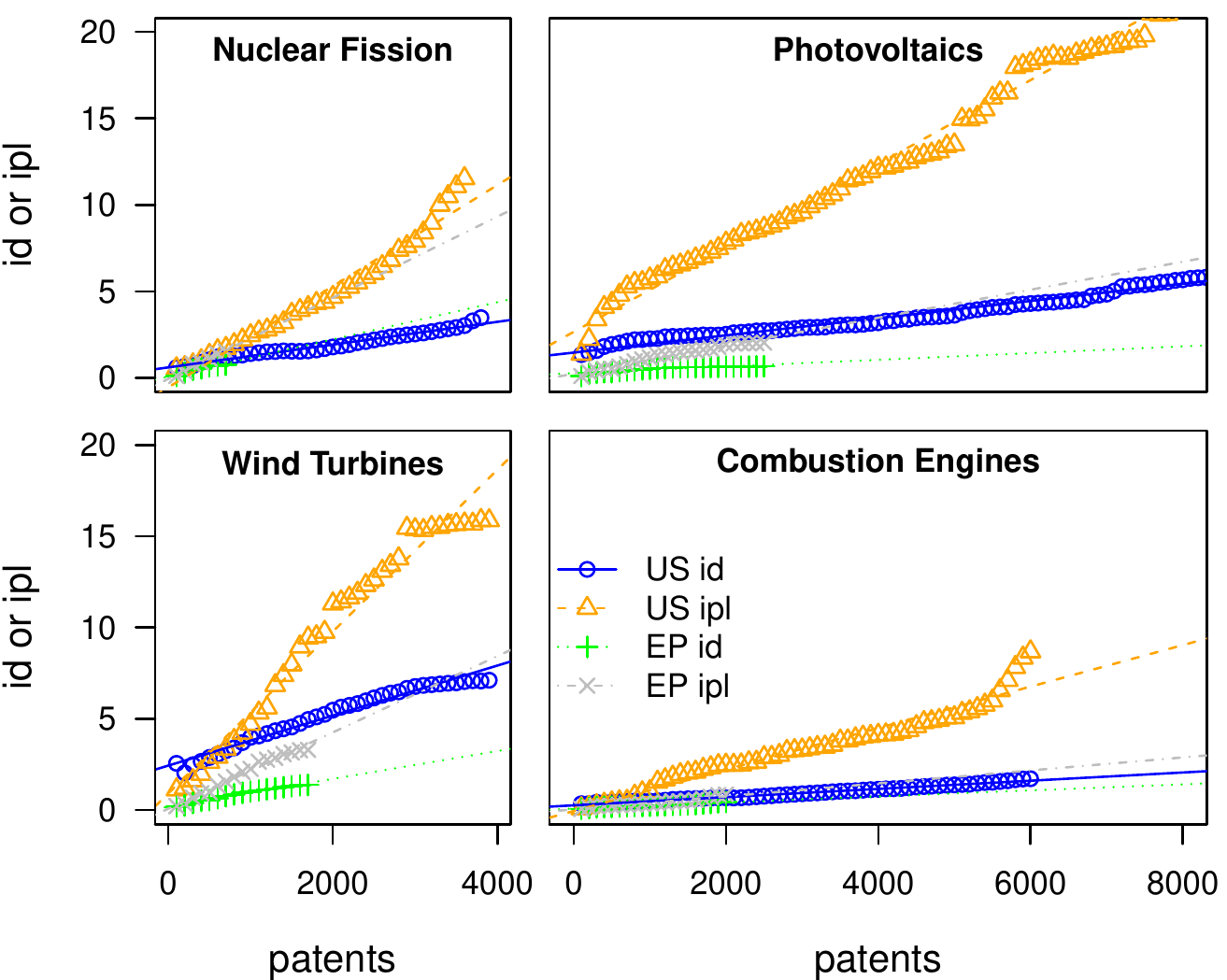}
\caption{\textbf{Id and ipl for number of patents (US and European patents)} We plot the id and ipl for every 100 patents (represented by symbols) and linear fits (represented by lines). For statistical details of the fit see Appendix \ref{appendix_3}.}
\label{id_ipl_patents}
\end{wrapfigure}

In Figure \ref{id_ipl_patents} we plot the id and ipl of the four focus technologies for the number of patents. We include the results from both the US and EP patents. We observe for all four technologies a linear increase of both the id and ipl, yet the rate of increase varies considerably across technologies. In the US data set, where wind turbines is after 2000 patents already at an ipl of 10, combustion engines reaches the same ipl only after 6000 patents. These variations are also found considering the id or EP dataset instead. It is therefore instructive to consider not only the absolute cumulativeness of a technology, but also its cumulativeness relative to the size of its knowledge base. 

To obtain a more detailed understanding of the linear relationship between cumulativeness and the number of inventions, we consider the coefficients of the linear fits in Figure \ref{id_ipl_patents} for US patents in Table \ref{coefficients} and for the EP patents in Table \ref{coefficients2}, the statistical details of these fits can be found in Appendix \ref{appendix_3}. The coefficients in Table \ref{coefficients} indeed vary considerably across technologies, and high values for $m_0$ correspond to high values of $q$. This suggest that if the need for specialized knowledge is high at initial stages of a technology, it also increases faster as the technology develops. More importantly, Table \ref{coefficients} shows that the fitted ipl coefficients $p$ (on the left, determined empirically) are in reasonable agreement with the predicted values (on the right, calculated). This suggests that the id and ipl are interrelated in accordance with the simple model described in section \ref{ipl_section}.  
\begin{table}[h!]
\centering
\caption{
\textbf{Coefficients of id and ipl for US.} On the left we present the fitted id and ipl coefficients and the id constants from Figure \ref{id_and_ipl} for US patents. On the right we present the predicted ipl coefficients, where $p'_a$ is directly based on the id data and is $p'_b$ calculated using the fitted $m_0$ and $q$ (see Section \ref{ipl_section}). With the exception of US wind turbines, these predictions agree rather well with the fitted ipl coefficients. As expected $p'_a$ is generally more accurate than $p'_b$. For statistical details see Appendix \ref{appendix_3}}
\scriptsize
\begin{tabular}{lllllll}
\multicolumn{1}{l|}{} &
  \multicolumn{1}{l|}{\begin{tabular}[c]{@{}l@{}}id const\\ $m_0$ (ref/pat) \end{tabular}} &
  \multicolumn{1}{l|}{\begin{tabular}[c]{@{}l@{}}id coeff\\ $q$ (ref/pat$^2$)\end{tabular}} &
  \multicolumn{1}{l|}{\begin{tabular}[c]{@{}l@{}}ipl coeff\\ $p$ (1/pat)\end{tabular}} &
  \multicolumn{1}{l|}{} &
 \multicolumn{1}{l|}{$p'_a$ (1/pat)} &
  \multicolumn{1}{l|}{$p'_b$ (1/pat)} \\ \cline{1-4} \cline{6-7} 
\multicolumn{1}{l|}{Nuclear Fission} &
  \multicolumn{1}{l|}{0.65} &
  \multicolumn{1}{l|}{0.0006} &
  \multicolumn{1}{l|}{0.0029} &
  \multicolumn{1}{l|}{} &
  \multicolumn{1}{l|}{0.0029} &
  \multicolumn{1}{l|}{0.0022} \\ \cline{1-4} \cline{6-7} 
\multicolumn{1}{l|}{Photovoltaics} &
  \multicolumn{1}{l|}{1.45} &
  \multicolumn{1}{l|}{0.0005} &
  \multicolumn{1}{l|}{0.0024} &
  \multicolumn{1}{l|}{} &
  \multicolumn{1}{l|}{0.0024} &
  \multicolumn{1}{l|}{0.0020} \\ \cline{1-4} \cline{6-7} 
\multicolumn{1}{l|}{Wind Turbines} &
  \multicolumn{1}{l|}{2.42} &
  \multicolumn{1}{l|}{0.0014} &
  \multicolumn{1}{l|}{0.0044} &
  \multicolumn{1}{l|}{} &
  \multicolumn{1}{l|}{0.0067} &
  \multicolumn{1}{l|}{0.0067} \\ \cline{1-4} \cline{6-7} 
\multicolumn{1}{l|}{Combustion Engines} &
  \multicolumn{1}{l|}{0.26} &
  \multicolumn{1}{l|}{0.0002} &
  \multicolumn{1}{l|}{0.0011} &
  \multicolumn{1}{l|}{} &
  \multicolumn{1}{l|}{0.0011} &
  \multicolumn{1}{l|}{0.0006} \\ \cline{1-4} \cline{6-7} 
  \end{tabular}
\label{coefficients}
\end{table}
\begin{table}[h!]
\centering
\caption{
\textbf{Coefficients of id and ipl for EP.} Same as Table \ref{coefficients}, but then for EP patents.}
\scriptsize
\begin{tabular}{lllllll}
\multicolumn{1}{l|}{} &
  \multicolumn{1}{l|}{\begin{tabular}[c]{@{}l@{}}id const\\ $m_0$ (ref/pat) \end{tabular}} &
  \multicolumn{1}{l|}{\begin{tabular}[c]{@{}l@{}}id coeff\\ $q$ (ref/pat$^2$)\end{tabular}} &
  \multicolumn{1}{l|}{\begin{tabular}[c]{@{}l@{}}ipl coeff\\ $p$ (1/pat)\end{tabular}} &
  \multicolumn{1}{l|}{} &
  \multicolumn{1}{l|}{$p'_a$ (1/pat)} &
  \multicolumn{1}{l|}{$p'_b$ (1/pat)} \\ \cline{1-4} \cline{6-7} 
\multicolumn{1}{l|}{Nuclear Fission} &
  \multicolumn{1}{l|}{0.07} &
  \multicolumn{1}{l|}{0.0011} &
  \multicolumn{1}{l|}{0.0023} &
  \multicolumn{1}{l|}{} &
  \multicolumn{1}{l|}{0.0024} &
  \multicolumn{1}{l|}{0.0017} \\ \cline{1-4} \cline{6-7} 
\multicolumn{1}{l|}{Photovoltaics} &
  \multicolumn{1}{l|}{0.25} &
  \multicolumn{1}{l|}{0.0002} &
  \multicolumn{1}{l|}{0.0008} &
  \multicolumn{1}{l|}{} &
  \multicolumn{1}{l|}{0.0008} &
  \multicolumn{1}{l|}{0.0010} \\ \cline{1-4} \cline{6-7} 
\multicolumn{1}{l|}{Wind Turbines} &
  \multicolumn{1}{l|}{0.18} &
  \multicolumn{1}{l|}{0.0008} &
  \multicolumn{1}{l|}{0.0021} &
  \multicolumn{1}{l|}{} &
  \multicolumn{1}{l|}{0.0019} &
  \multicolumn{1}{l|}{0.0016} \\ \cline{1-4} \cline{6-7} 
\multicolumn{1}{l|}{Combustion Engines} &
  \multicolumn{1}{l|}{0.07} &
  \multicolumn{1}{l|}{0.0002} &
  \multicolumn{1}{l|}{0.0004} &
  \multicolumn{1}{l|}{} &
  \multicolumn{1}{l|}{0.0005} &
  \multicolumn{1}{l|}{0.0005} \\ \cline{1-4} \cline{6-7} 
\end{tabular}
\label{coefficients2}
\end{table}
This implies that the relation between id and ipl is rather predictable for each technology, suggesting that the transversal and longitudinal dimensions of cumulativeness (using a proper re-scaling) can be used interchangeably. 
In Table \ref{coefficients2} the variation across technologies for EP patents is largely similar to the variation for US patents, and again shows reasonable agreement between the fitted and predicted ipl coefficients. There are however also some overall differences with the US patents. The constants $m_{0}$ are generally smaller and, as a consequence, the ipl coefficients $p$ are also smaller. There are minor differences per technology, the id coefficient $q$ of nuclear fission being remarkably higher for the EP patents than for the US patents. We will revisit cross technology differences more systematically at the end of this chapter. 

\begin{wrapfigure}{r}{0.6\textwidth} 
\centering
\includegraphics[width=0.6\textwidth, height=0.5\textwidth]{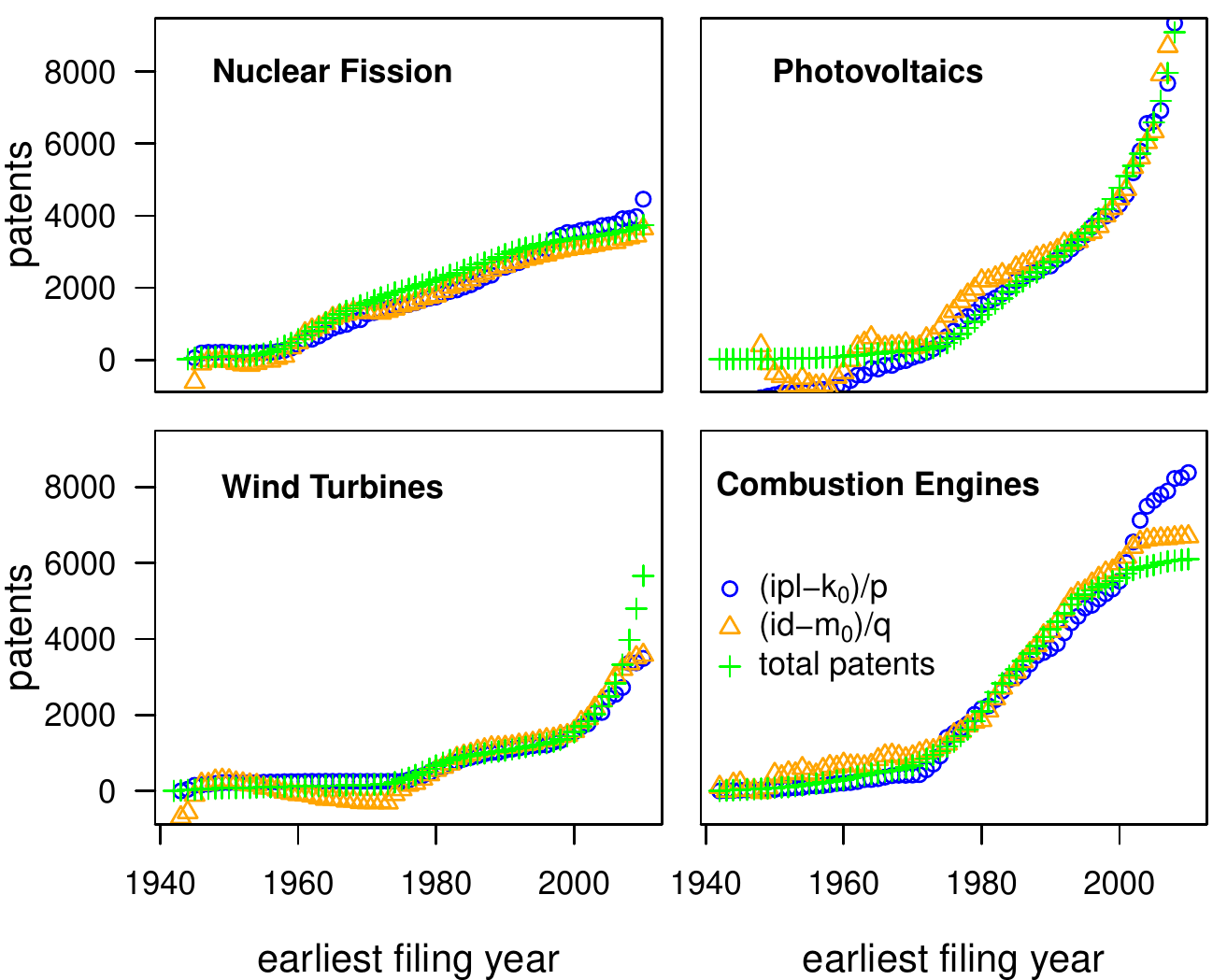}
\caption{\textbf{Total patents and re-scaled id and ipl over time} For each earliest filing year we plot the total number of patents and the ipl and id, where ipl-$k_{0}$ is re-scaled by a factor $1/p$ and id-$m_{0}$ is re-scaled by a factor $1/q$ (the factors are taken from Table \ref{coefficients} and appendix \ref{appendix_3}). Both cumulativeness indicators closely follow the development of the total patents over time.}
\label{id_ipl_time}
\end{wrapfigure}  

Finally we observe in Figure \ref{id_ipl_patents} some minor deviations from the linear developments, in particular the ipl of nuclear fission and combustion engines speeding up for higher number of patents, and that of wind turbines slowing down. Additionally, the ipl of combustion engines and photovoltaics increases fast at lower number of patents. (A closer analysis of the id leads to similar observations, though this is less clear in Figure \ref{id_ipl_patents}). We will come back to these deviations in our discussion of Figure \ref{id_ipl_time}. 

In Figure \ref{id_ipl_time} we plot for the US patents the id and ipl over time, together with the total number of patents over time.\footnote{The number of backward citations only starts to become substantial from 1940 onward for all considered technologies, which is why we choose this as a starting point. We note that wind turbines have a substantial a number of patents (about 1300) before 1940, yet citations before that period are either rare or not recorded in our data set.} The ipl values (shifted by $k_{0}$) are re-scaled by the corresponding factor $p$ and the id values (shifted by $m_{0}$) are re-scaled by the corresponding factor $q$ from Table \ref{coefficients}. 
We observe for all four technologies that the time development of all three quantities largely coincide. In hindsight, this should not be a surprise given the observed linear relations in Figure \ref{id_ipl_patents}: the id and ipl are mainly a function of the total number of patents and hence their developments are synchronized. The synchronization indicates that our modeling of the knowledge dynamics consistently applies over time, i.e. that it is to some extent time independent. Still, we note the synchronization is not always perfect: towards 2009, we observe that the ipl of nuclear fission and especially combustion engines increases faster than the number of patents, and vice versa for wind turbines. Further, in the 1960s, the ipl of photovoltaics and combustion engines is somewhat lower than the number of patents. Note that these asynchronous developments correspond exactly to the previously mentioned deviations from linearity in Figure \ref{id_ipl_patents}. Note in Figure \ref{id_ipl_time} that the 'fast ipl' deviations correspond to periods in time where the number of patents increase very slow, (nuclear fission and combustion engines towards 2009, photovoltaics and combustion engines in the 1960s) and that the 'slow ipl' deviations correspond to periods in time where the number of patents increase very fast (wind turbines towards 2009). To some extent, but less clearly in Figure \ref{id_ipl_time}, this also counts for the id developments. These observations are therefore in agreement with the hypothesized inverse relation between the rate of invention over time and cumulativeness coefficients.  

\subsection{Distributions of backward links and path length} 

The measured linear relationships between the id, ipl and number of patents are in line with the model predictions of Section \ref{dynamics}, yet linear relationships may also arise in various other models. Additionally we therefore study the empirical backward link and path length distributions and compare these to the predicted distributions. For brevity we focus on the US patents in this section, as the analysis for EP patents is largely similar. 
\\

\begin{figure}
\centering
\begin{minipage}{.48\linewidth}
  \includegraphics[width=\linewidth]{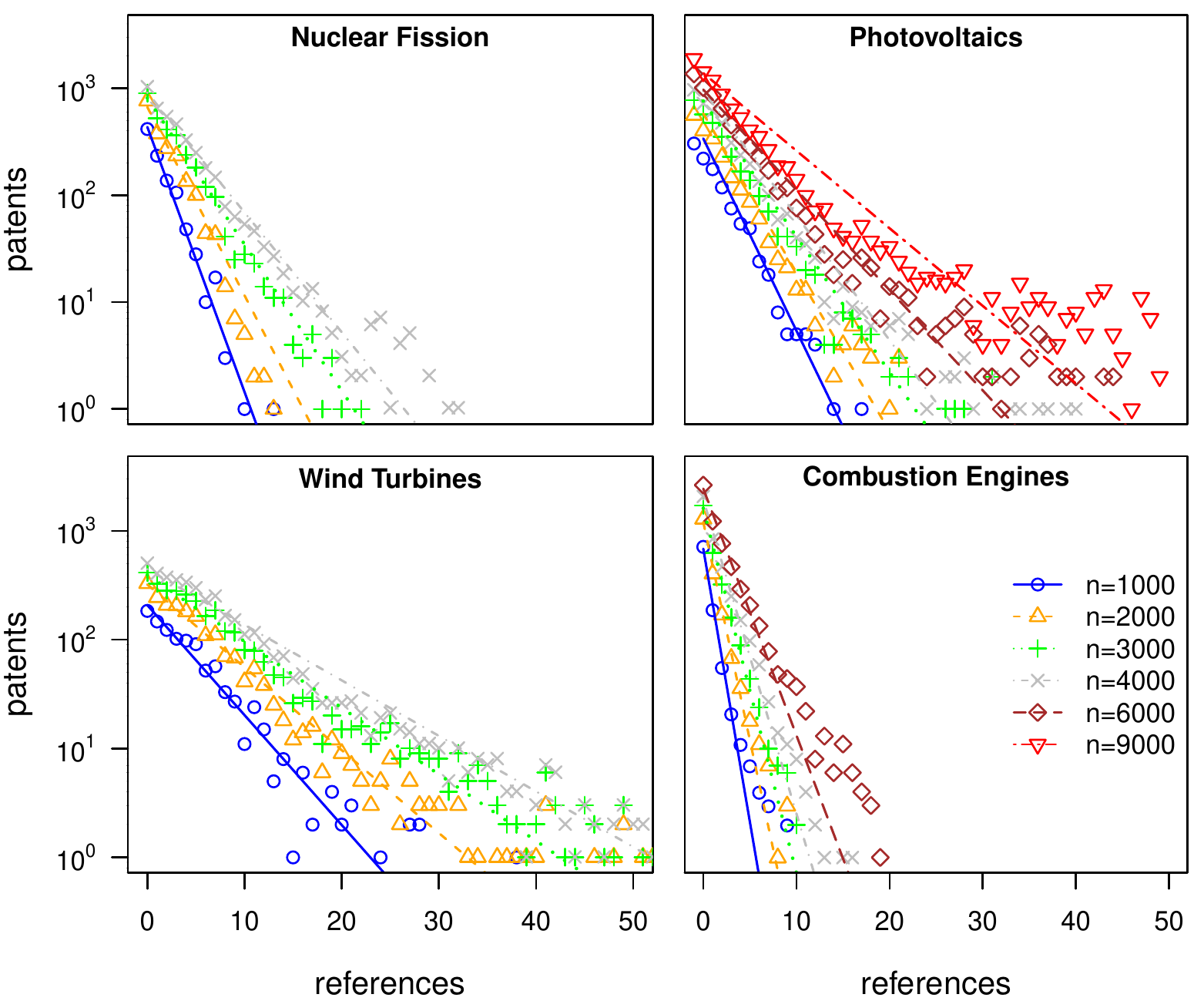}
  \caption{\textbf{Backward link distribution (US patents)} With symbols we plot the empirical distribution each time the number of patents increase by a 1000. With lines we plot the predicted geometric distributions using the parameters $q$ and $m_{0}$ from Table \ref{coefficients}. For clarity, the $n=5000$,$7000$ and $8000$ are omitted (applicable to combustion engines and photovoltaics only).}
\label{id_dis}
\end{minipage}
\hspace{.02\linewidth}
\begin{minipage}{.48\linewidth}
  \includegraphics[width=\linewidth]{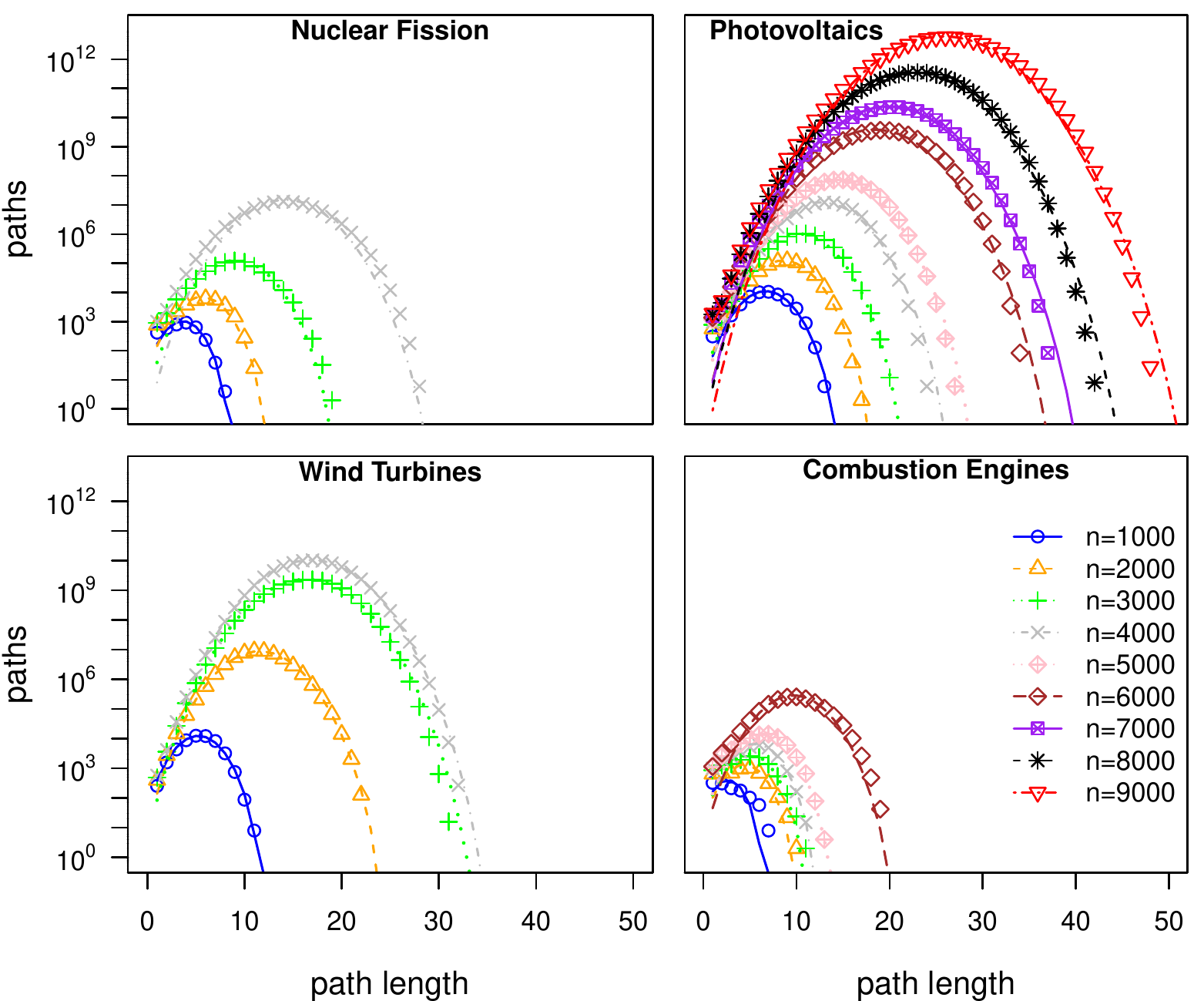}
  \caption{\textbf{Path length distribution (US patents)} With symbols we plot the empirical distribution each time the number of patents increase by a 1000. With lines we plot the predicted distributions from Equation \ref{main7}, where the maximum path length values plotted by numerals in Figure \ref{max_speed} are used as values of $n'$.
  \\ }
\label{ipl_dis}
\end{minipage}
\end{figure} 
In Figure \ref{id_dis} we plot the internal backward link distribution for the four focus technologies for US patents, plotting the distribution for each technology for every 1000 patents. We observe two characteristics: (1) the frequency drops exponentially (note the logarithmic axis) with the number of references and zero references occuring most frequently, (2) as the number of patents increase, the skewness decreases. Where (1) is indicative for a geometric distribution, (2) indicates that the parameter of this distribution depends on the number of patents. To test if these distributions agree with the predictions of Section \ref{id_section}, we in Figure \ref{id_dis} simultaneously plot the predicted distributions using the parameters $q$ and $m_0$ from Table \ref{coefficients}. We observe the predicted distributions fit the empirical distributions rather well. In appendix \ref{appendix_4} we compare these fits to a number of alternative distributions using probability plots, which again confirm the data is reasonably well described by geometric distributions with parameters from Table \ref{coefficients}. 
 
\begin{wrapfigure}{l}{0 pt} 
\centering
\includegraphics[width=0.5\textwidth, height=0.3\textwidth]{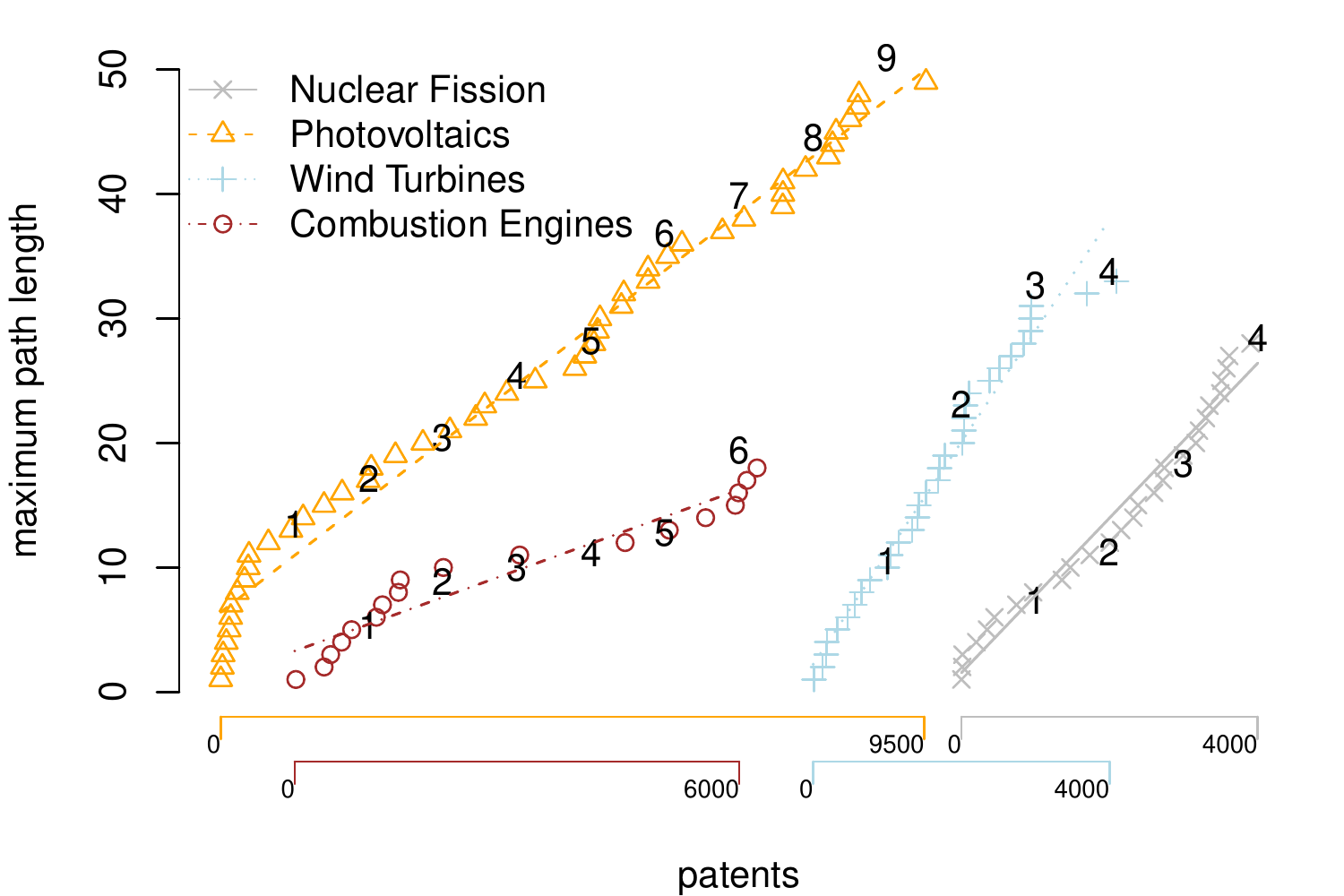}
\caption{\textbf{Maximum path length for US patents} We plot the maximum internal path length for the number of patents (in symbols) and include linear fits (in lines) of the development. Details of the fits are included in Appendix \ref{appendix_3}. The plotted numerals each correspond to one of the fitting distributions plotted in Figure \ref{ipl_dis}, the x-coordinate representing the number of patents $n$, the y-coordinate representing the $n'$ used in the fitting distribution (see equation \ref{main7}).}
\label{max_speed}
\end{wrapfigure}

In Figure \ref{ipl_dis} we consider the path length distribution (for each internal path) for US patents, plotting the distribution for every 1000 patents. We observe two characteristics: (1) as the number of patents increase, the distribution becomes less skewed, approximating a parabolic shape (on a log scale) (2) the most frequent path length shifts right as the number of patents increase. 
Before we discuss the fitting of the path length distributions, we shortly consider the development of the maximum internal path length (mipl) for the number patents in Figure \ref{max_speed}. The patterns are for each technology rather similar to those of the ipl in Figure \ref{id_ipl_patents}, except that the mipl increases at about double the pace. 

Indeed, linear fits of the mipl, see Appendix \ref{appendix_3} for the details, yield the coefficients 0.0062 (nuclear fission), 0.0046 (photovoltaics), 0.0089 (wind turbines) and 0.0021 (combustion engines), which are as expected very close to $2p$ (using the values $p$ from Table \ref{coefficients}). As explained at the end of Section \ref{ipl_section}, we can use the maximum path lengths as estimates for $n'$ in Equation \ref{main7}. The values of $n'$ used in the fitting path length distributions in Figure \ref{ipl_dis} are the y-coordinates of the plotted numerals in Figure \ref{max_speed}. We note that the empirical distributions in Figure \ref{ipl_dis} are very well fitted, and at the same time the numerals in Figure \ref{max_speed} fall neatly in the line of development of each technology. We therefore conclude that Equation \ref{main7} provides a rather accurate description of the path length distributions. A closer examination of the distribution fits is provided in Appendix \ref{appendix_4}. 

\subsection{Cross technology variations}\label{crosstech}
\begin{figure}
\centering
\caption{\textbf{Id and ipl for 24 technologies in EPO and USPTO.} Linear fits are included based on the pairwise regressions in Appendix \ref{appendix_3}.}
\includegraphics[width=\textwidth]{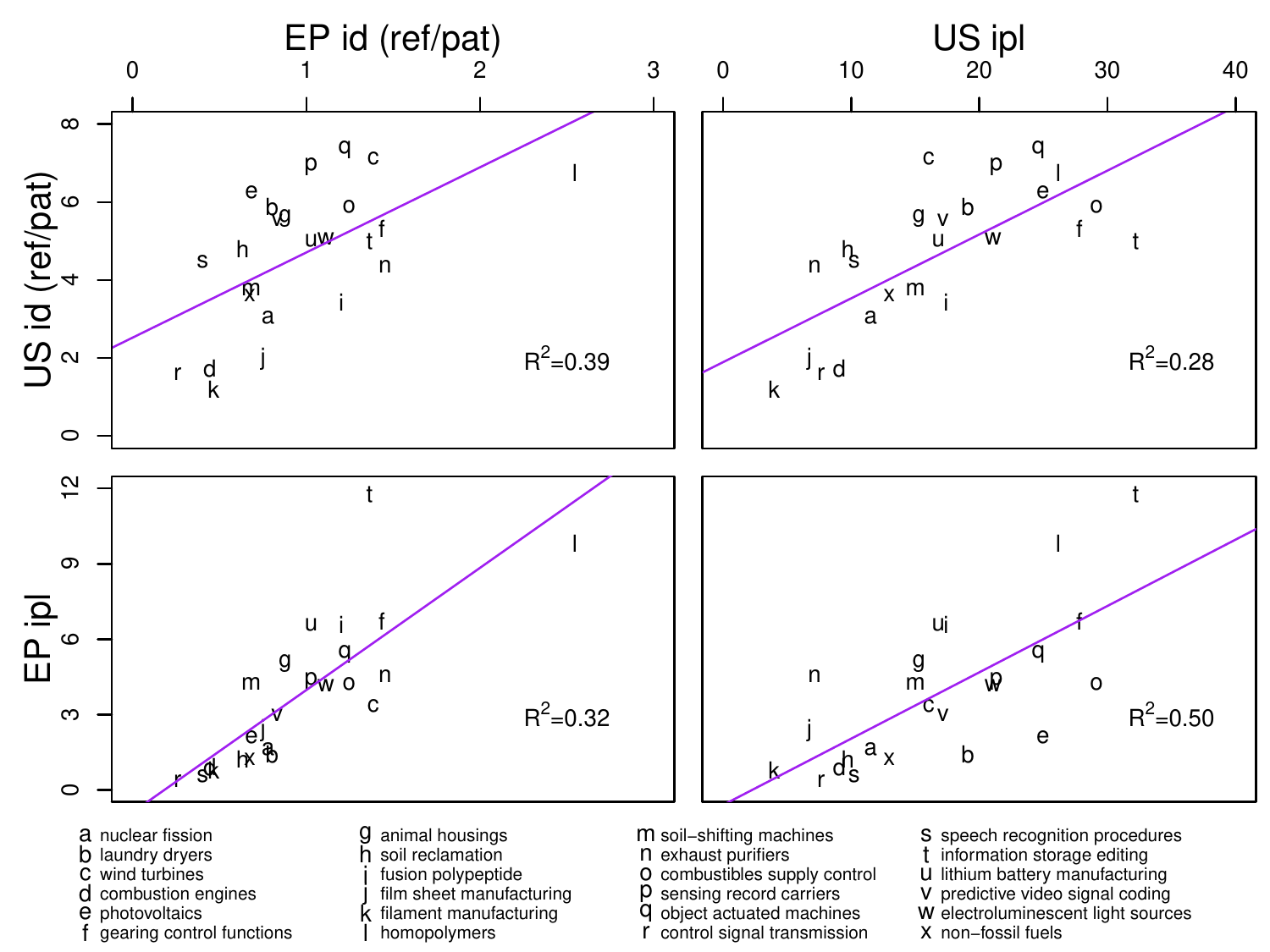}
\label{PL_IKD}
\end{figure}

Finally we discuss the variation in id and ipl across technologies in more detail. As we have two indicators for cumulativity and two data sources (EPO and USPTO), we can identify cross cumulativity variations along 4 dimensions. Figure \ref{PL_IKD} presents a systematic overview of the 24 technologies from Table \ref{desstat} and \ref{addtech}. We observe positive trends for each comparison in Figure \ref{PL_IKD}, and the technologies for each comparison remain rather consistently in a characteristic (high or low) range of cumulativity. These observations are supported by the reasonable values of the squared correlation coefficient $R^{2}$ and the statistical significance we find for each comparison (for the statistical details see Appendix \ref{appendix_3}). The positive association between the two indicators in Figure \ref{PL_IKD} provides some evidence that the relation established in Equation \ref{slope} applies across a wider range of technologies than the four focus technologies. This suggests again that the degree of cumulativeness measured along the transversal and longitudinal dimensions largely agree for each technology, i.e. that both dimensions can be used more or less interchangeably. As expected from the greater citation tendency in the US system, the values of the US indicators are a factor 3-4 greater than their EP counterparts (see also Appendix \ref{appendix_3}). However, the positive association we find between US and EP patents for both the id and ipl indicates that this factor is approximately constant across different technologies. This suggest that, despite institutional differences, both indicators can be applied consistently within different patent systems, confirming that we can think of cumulativity as a technology specific characteristic. 

\begin{figure}
\centering
\caption{\textbf{Invention rate and id rate $q$} The symbols correspond to the technologies in the legend of Figure \ref{PL_IKD}. Note the axes are logarithmic, hence the fitted line is a power-law (for details see Appendix \ref{appendix_3}).}
\includegraphics[width=\textwidth]{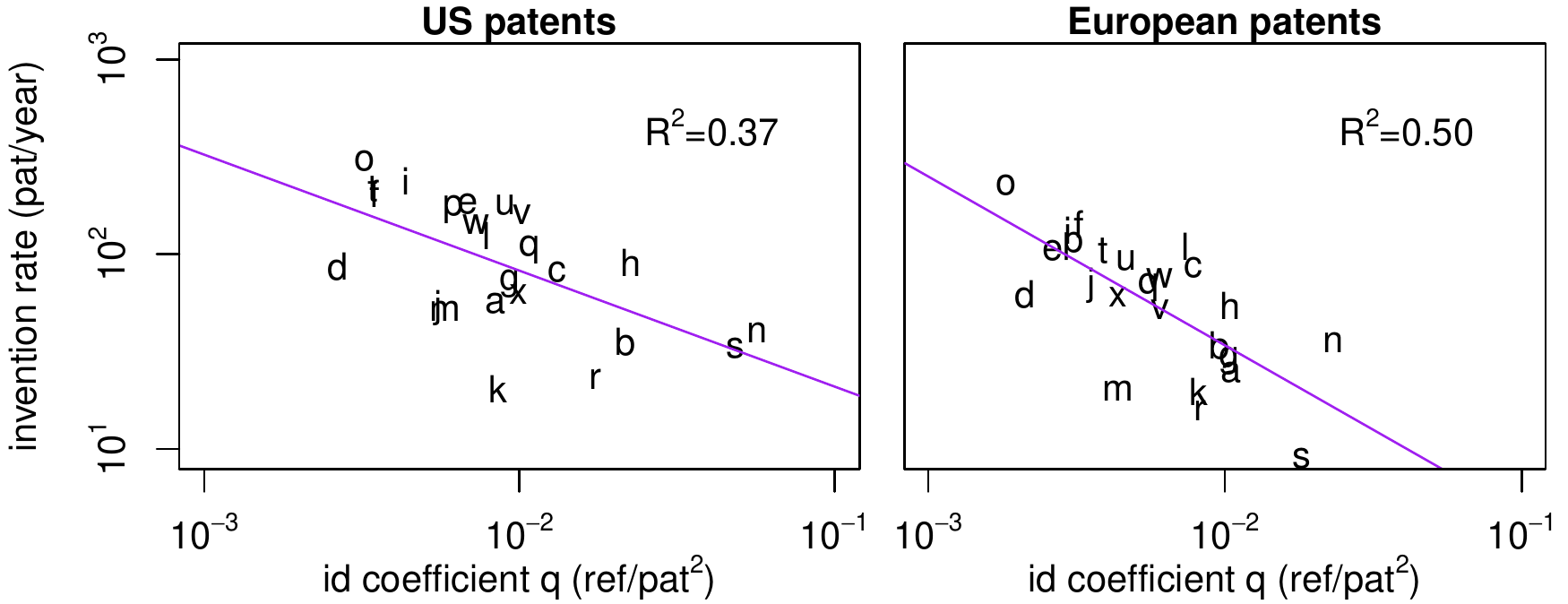}
\label{rate_time}
\end{figure}

In our discussion of the four focus technologies we provided some evidence for the hypothesized inverse relation between the time rate of invention and the id coefficient $q$ (i.e. the rate at which the id increases per patent). The joint consideration of 24 technologies allows us to test this relation for a wider range of technologies. In Figure \ref{rate_time} the invention rate (measured by the average number of patents per year) is plotted for the id coefficients $q$ (determined by the number of references per patent squared) for both the US and EP patents. In line with expectation, the two quantities are negatively associated (best fitted by a power law with a power $\approx -0.6$ for US patents and $\approx -0.9$ for EP patents). Again see Appendix \ref{appendix_3} for the statistical details. Figure \ref{rate_time} therefore confirms that the linear coefficient determining the increase of the id per patent (and indirectly the ipl) is related inversely with the rate of invention over time. Note that this does not mean the rate of cumulativeness development is exclusively determined by the rate of invention, as there may still be other factors at play related to the type of technology or technological knowledge. From Figure \ref{PL_IKD} it is not directly clear however what type of technologies we can typically associate with high and low cumulativeness: we observe technologies from various disciplines both on the higher and lower end of the spectra. In the final subsection we therefore consider the differences between technologies on a more aggregated level of classification. 
\subsection{Cumulativeness across technological fields}
\begin{figure}
\centering
\caption{\textbf{Cumulativeness versus size of knowledge base for US patents} We plot the cumulativeness (measured by the internal dependence) for the knowledge base size (measured by the number of patents) for 40 technological fields based on USPTO data. Fields in the same CPC section are colored similarly. Note both axes are logarithmic, hence the fitted regression line is a power law. The cumulativeness of technologies appearing substantially above (below) the fitted line can identified as relatively high (low).}
\includegraphics[width=\textwidth]{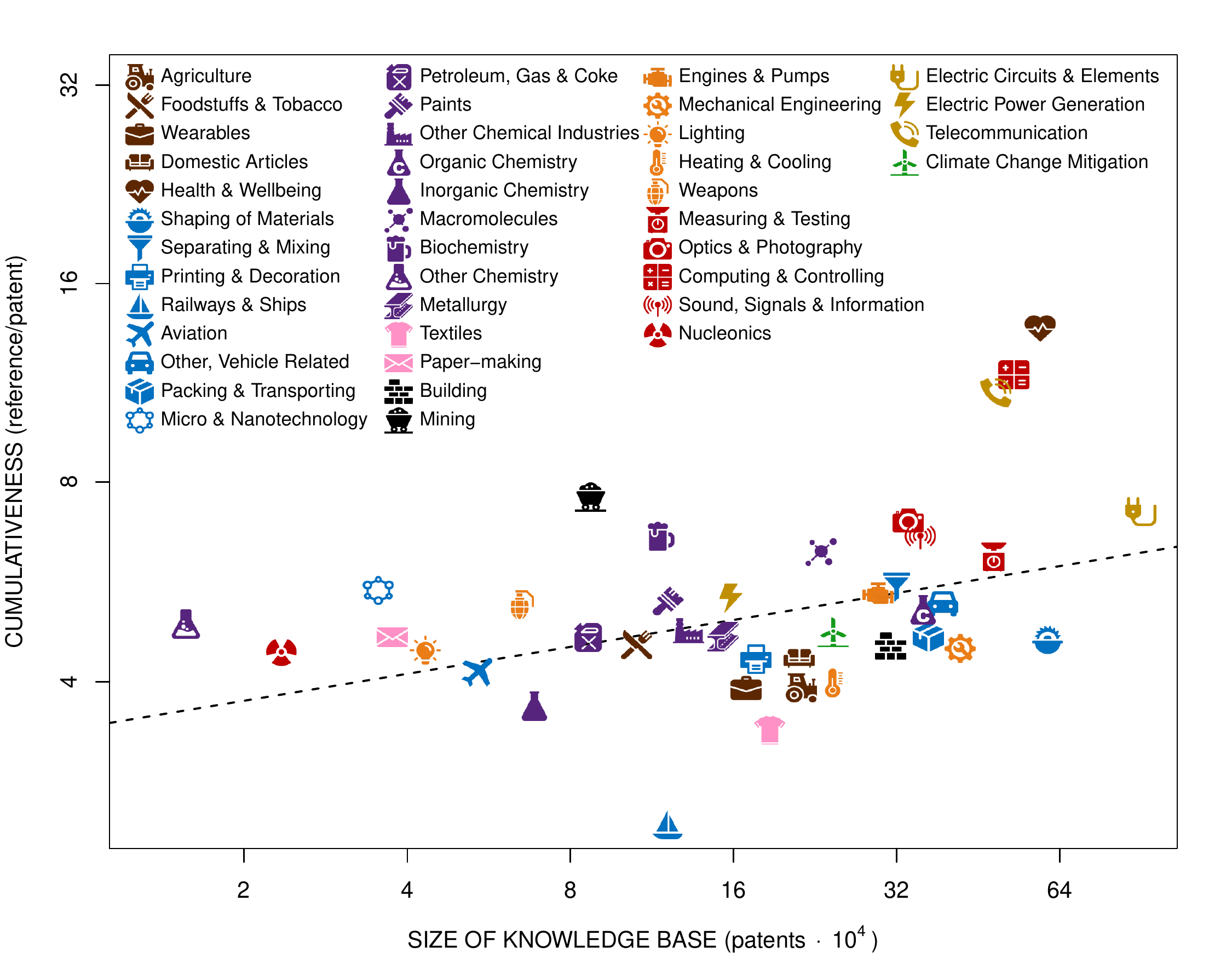}
\label{icon_US}
\end{figure}

\begin{figure}
\centering
\caption{\textbf{Cumulativeness versus size of knowledge base for European patents} Same as Figure \ref{icon_US} but then for European patents. A legend for the icons is included in Figure \ref{icon_US}.}
\includegraphics[width=\textwidth]{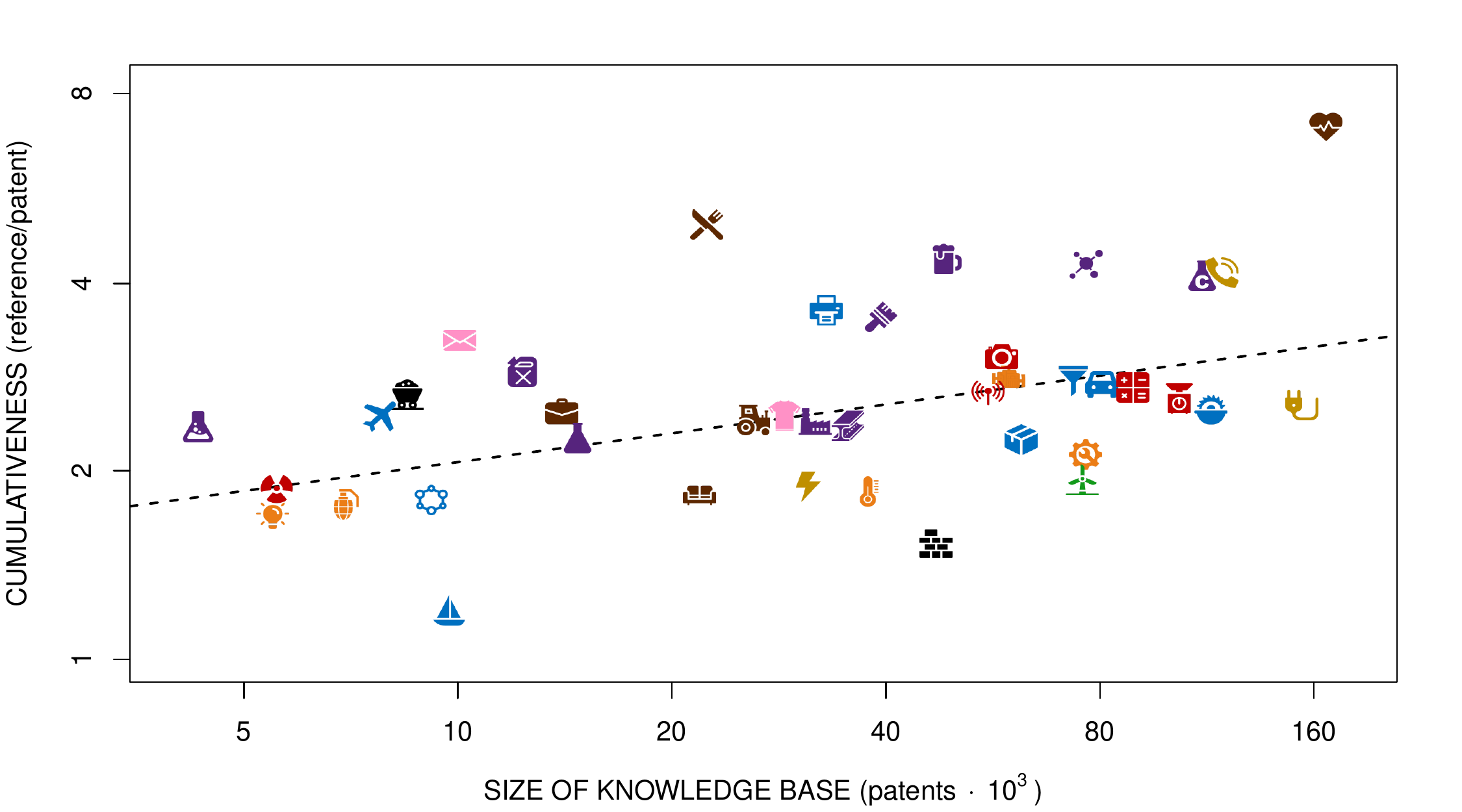}
\label{icon_EP}
\end{figure}

Finally we consider the cumulativeness of technologies on a more aggregated classification level, which we will henceforth refer to as 'technological fields', (for an overview see Table \ref{super_class} in Appendix \ref{appendix_3}). This allows us to develop an overall understanding of which technologies can typically be associated with high or low cumulativeness. Furthermore, it allows us to check if our approach to cumulativeness is in line with earlier approaches \parencite{malerba_schumpeterian_1996,breschi_technological_2000}\footnote{The contribution by Breschi is largely consistent with the one by Marlerba and Orsenigo. As the latter considers more detailed technological classes and a wider geographical range of patents, we will focus on the latter.}, thus to some extent validating the indicators for cumulativeness suggested this contribution. However, as determining the ipl is computationally challenging for very large numbers of nodes (i.e. >100,000), we limit this analysis to determining the id of these technological fields. We plot the id for the number of patents for these fields for the US patents in Figure \ref{icon_US}, where we also include a legend. Note that the different icon colors correspond to the different CPC main sections. Figure \ref{icon_EP} shows a similar plot for the European patents. 

For a deeper understanding of a technology's cumulativeness, we again stress the need to additionally consider the cumulativeness relative to the size of the knowledge base. For example, in Figure \ref{icon_US}, while the knowledge base size is similar for the field Packing \& Transporting and the field Optics \& Photography, the latter has reached a far greater level of cumulativeness. Similarly, Nucleonics reaches the same cumulativeness level as Packing \& Transporting while the knowledge base is about 15 times larger in the latter. The cumulativeness therefore appears to increase faster with each patent for Nucleonics and Optics \& Photography than for Packing \& Transporting. The expected increase in cumulativeness for the knowledge base size is indicated by the fits (dashed line) in Figures \ref{icon_US} and \ref{icon_EP} and may depend on the level of classification. 
For this level of classification we can use these fits to distinguish \textit{relatively high} cumulativeness (above the line) from \textit{relatively low} cumulativeness (below the line). Using this distinction we see for the US patents the fields belonging to CPC sections Physics (red icons), Electricity (yellow icons) and Chemistry (purple icons) show relatively high levels of cumulativeness. Fields belonging to Sections Human Necessities (brown icons) and Performing operations \& Transporting (blue icons) show relatively low levels of cumulativeness. The larger fields in the Sections Textiles (pink icons) and Fixed Constructions (black icons) too show relatively low levels of cumulativeness. 

The study by Malerba and Orsenigo (M\&O) distinguishes a number of highly aggregated technologies as Schumpeter Mark I (associated with low cumulativeness) and Schumpeter Mark II (associated with high cumulativeness). Our observations are in overall agreement with the general conclusion of M\&O that \textit{"Schumpeter Mark I technological classes are to be found especially in the 'traditional' sectors, in the mechanical technologies, in instruments as well as in the white electric industry. Conversely, most of the chemical and electronic technologies are characterized by the Schumpeter Mark II model."}\footnote{We interpret M\&Os 'traditional' sectors to correspond to the early industrial and craft-like sectors such as Textiles, Domestic Articles and Wearables}. To make a more detailed comparison, we individually consider 23 technological fields which occur both in the M\&O and our own set of fields and which M\&O classify as either Schumpeter I or II. For the purpose of this comparison, we associate a technological field below the fitted line with low cumulativeness (which should correspond to M\&O's Schumpeter Mark I) and technologies on or above with high cumulativeness (corresponding to M\&O's Schumpeter Mark II). From the 23 thus considered technologies, 18 are identified correctly: 7 as low cumulativeness (Wearables, Domestic Articles, Agriculture, Shaping of Materials, Railways \& Ships, Building, Mechanical Engineering) and 11 as high cumulativeness (Aviation, Petroleum, Gas \& Coke, Macromolecules, Biochemistry, Engines \& Pumps, Weapons, Photography \& Optics, Nucleonics, Telecommunications, Computing \& Controlling, Electronic Components \& Circuitry). 5 Technological fields do not correspond to M\&O's labeling: Inorganic Chemistry (Mark II), Printing \& Decoration (Mark II), Lighting (Mark I), Measurement \& Testing  (Mark I) and Health and Wellbeing (Mark II). Note the first four are rather close to the line however. The cumulativeness of Health and Wellbeing is exceptionally high though in our analysis. The reason for these deviations is not directly clear. We emphasize that M\&O's Schumpeter Mark I or II labels are based on various aspects of the organization of innovation, and are therefore only an indirect indication of cumulativeness. Also, there might be some variation between the grouping of patent classes by M\&O and ours. Finally, some technologies may have developed substantially between the M\&O study (1996) and the final year we consider (2009).  

The variations found across technological fields using the European patents in Figure \ref{icon_EP} are largely similar to those we observed for the US patents. Notable differences are that for the European patents, the chemistry fields show relatively high cumulativeness and the physics and electricity fields show relatively low cumulativeness (as compared to the US patents). In general the variations across technological fields are less for the European patents than for the US patents, which is likely related to the fact that the number of patents is substantially lower for the former. Although there are some differences with the US patents, the European patents too show overall agreement with the results of M\&O (of the 23 fields, 18 are identified correctly). The agreement between the M\&O approach to cumulativeness and our results provide a validation for the use of the id to measure the cumulativeness of a technology, and indirectly for the ipl, given the earlier established close relation between both indicators.   

\section{Discussion}
\label{Discussion}

In this paper we established an approach to interpret, model and measure the cumulative nature of technological knowledge development. We can identify an number of deeper implications and possible extensions of the theoretical model developed in this contribution. 

A main point in the search model is the increasing difficulty to invent without any prior knowledge of the field, which leads to a geometric distribution for the number of backward links. In a number of other approaches, invention is perceived as a process of (re)combining existing pieces knowledge \parencite{fleming_recombinant_2001, fleming_technology_2001, arthur_nature_2009}. When we would focus on the number of combinations allowed by the number of existing inventions, a reasonable suggestion for the distribution of backward links would be a binomial type of distribution. This option may seem attractive, as assigning equal probability to each combination would lead the expected value of the number of backward links per invention to increase proportionally with the number of inventions, in agreement with observation. However, for the id we fail to observe characteristics of a binomial type of distribution. The fact that we obtain stronger evidence in Appendix \ref{appendix_4} for a geometric distribution suggests therefore that the mechanism of combination plays a lesser role than we might expect, or at least that we are dealing with a special type of combination, where for example only a small subset of the combinations is allowed. 

While linear relations are common in descriptions of social phenomena, we emphasize that the linearity of the id and ipl in the number of inventions is neither an obvious nor an expected result. In a number of network approaches to knowledge dynamics it is instead supposed that the number of backward links per node is on average constant as the number of nodes increase \parencite{wang_quantifying_2013,albert_statistical_2002, price_general_1976}. It can be demonstrated this would imply a constant id and a logarithmically increasing or even constant ipl. These mechanisms would thus predict a stagnating cumulativeness, even though the number of inventions keeps increasing. One may raise the objection that the external nodes are not included in our analysis, and that the id linearity may disappear once these are included. Additional checks on the four focus technologies in Table \ref{desstat} however reveal that the external dependence (i.e. the average number of external nodes each node builds on) equally well shows a linear increase. Although considering only four technologies gives no guarantee, it is an indication that the linearity is a more general phenomenon. In this contribution we explored some possible mechanisms driving the increase of id and ipl. At the same time we acknowledge that there may be other societal factors driving the increase, such as increased computerization or other factors improving the availability of search results. Accounting for the effect of these factors is however challenging, as it would require us to compare similar technological developments over different time periods. 

Our approach suggests that the cumulativeness of technologies develops largely in sync with the size of the respective knowledge base, which suggests that these knowledge dynamics are to some extent time independent, i.e. less impacted by historical events. Likewise, the description was formulated independent of spatial (geographical) factors (and appeared consistent between the US and Europe). This appears to contradict a commonly held notion that technology development is highly path dependent, i.e. that history and local circumstances crucially matter. However, the time and space independence here only applies to the relation between cumulativeness and the size of the knowledge base, hence the crucial choices determining particular \textit{technological content} may still largely depend on historical or local events. Furthermore, we observed that the rate of invention of this technology over time is inversely related to the rate of proportionality between the cumulativeness and size of the knowledge base. If there is causation from former to latter, then technological cumulativeness may in the end be less determined by intrinsic knowledge properties than generally understood. If there is causation from latter to former, then the cumulativeness rate of a technology can be interpreted as key determinant and predictor of its rate of invention. Alternatively, a  simultaneous effect of both causalities may also be the case. Regardless of a possible causality direction, it would for later work be interesting to compare the deviations from linearity in Figure \ref{id_ipl_patents} with different phases in the technology life cycles \parencite{abernathy_patterns_1978, anderson_technological_1990}. The development of combustion engines and nuclear fission indeed show hints of typical life-cycle s-shapes in Figure \ref{id_ipl_time}, the points of acceleration and deceleration corresponding to the deviations in Figure \ref{id_ipl_patents}. While the present model does not account for these deviations, we note that, at least for the technologies here considered, the deviations are minor, and linearity remains the dominant pattern.

In this contribution we focused on the cumulativity of \textit{technological} knowledge. It would be interesting to compare this to cumulativity in other fields of knowledge such as science or art. The indicators and models discussed in this contribution can reasonably well be generalized to these areas. Also, it would be interesting to look at science-technology or art-technology dependencies, which then allow us to consider the cumulativity of technology as a whole, i.e. consider all technology as internal and the influence of science and/or art as 'external'. These questions are however beyond the scope of this work.

Finally we mention two limitations to our approach. First, our results critically depend on a particular choice for a demarcation/classification of different technologies, in our case the CPC. Even though a validated classification, innovation researchers should keep in mind the CPC is in the first place designed to aid patent examiners in their search for prior art, which may not always align with the technology definitions and level of detail researchers require. Furthermore, as new technologies develop the CPC is continuously restructured, causing possible misalignment with the researchers time perspective of a developing technology. To allow for a more detailed classification or a more sophisticated internal-external distinction researchers may consider alternatives based on textual analysis of patents \parencite{kelly_measuring_2018}, technological relatedness \parencite{castaldi_related_2015} or distance measures \parencite{gilsing_network_2008, jaffe_characterizing_1989}. While we acknowledge these points, we note that the main focus of this work was on developing a methodology to determine a technology's cumulativeness, which is generally applicable once the internal-external distinction is in place. In general we emphasize that a better understanding of the applicability of our analysis requires us to research a greater number of technologies. This would also help us understand if more closely related technologies also differ less in cumulativeness (hints of which we observe in Figures \ref{icon_US} and \ref{icon_EP}). Second, we kept the models in this contribution as simple as possible, thereby excluding a number of arguably relevant factors, amongst others: (i) the average time lag between the appearance of knowledge and the usage of that knowledge (ii) more advanced mechanisms in patent networks such as preferential attachment effect \parencite{albert_topology_2000,erdi_prediction_2013,valverde_topology_2007}, (iii) linkage to external inventions, which allows paths to start directly from external nodes. Though we can think of possible extensions of the model including these factors, we preferred a simple version for clarity.  

\section{Conclusions}
\label{conclusions} 

This paper presents both a theoretical and an empirical investigation of technological cumulativeness. Theoretical perspectives agree that technological cumulativeness involves a series of developmental steps within a technology, where the cumulativeness is higher (i) when the dependence between subsequent steps is larger, and (ii) when the total number of subsequent steps is higher. We capture these transversal (i) and longitudinal (ii) dimensions of cumulativeness through our indicators \textit{internal dependence} (id) and \textit{internal path length} (ipl).  

We then analytically derive how the id and ipl interrelate, and how they change as the size of the knowledge base of a technology increases (as measured by the total number of inventions). To this end, we model the invention process as a series of searches. A relevant parameter in this process is the technology specific rate $q$ at which it becomes harder to invent without using the existing knowledge in the field. We expect $q$ to be inversely related to the rate of invention over time, as there tends to be more specialization (and hence less need for complete knowledge) at greater rates of invention. From this model we deduce that the id and ipl, while following different distributions, are both expected to increase linearly with the size of the knowledge base. The coefficients of this linear relations are predicted to approximate $q$ as the knowledge base becomes larger.  

Empirical tests on several technologies, using patent and citation data from both USPTO and EPO as proxies for invention and knowledge flow, provide empirical support for these expectations and show that the id and ipl can be used consistently for both patent systems. Further, the variations in cumulativeness across technological fields are found to be largely consistent with earlier contributions which used different approaches to technological cumulativeness: chemistry, physics and to some extent electronics are generally characterized by relatively high cumulativeness, while the craft-like and mechanical engineering fields show relatively low cumulativeness.    

Our study leads to a number of new insights about technological cumulativity and its relation to technological knowledge: 
\begin{enumerate}
    \item The cumulativeness of a technology develops proportionally with the size of its knowledge base, with a technology specific \textit{cumulativeness rate}. A thorough understanding of a technology's cumulativeness therefore considers the cumulativeness both absolute as well as relative to the size of its knowledge base.  
    \item The measurements of cumulativity along the transversal dimension and the longitudinal dimension are found to be consistent for various technologies. It appears therefore that both provide an equivalent description of a technology's cumulativeness. Measuring the transversal dimension by means of the internal dependence is (computationally) simple, and therefore provides a relatively fast and reliable indication of a technology's cumulativeness.   
    \item The time development of the cumulativeness indicators is largely synchronized with the time development of the knowledge base size. This suggests that short term, immediate effects have a limited influence on the relation between cumulativeness and knowledge base size (meaning that the cumulativeness rate remains constant). However, across technologies we observe an inverse relation between the cumulativeness rate and the rate of invention over time. This suggests that effects acting over longer periods of time, such as the gradual acceleration or deceleration of inventive efforts, may therefore affect the cumulativeness rate.  
    \item Technological cumulativeness is understood to be a mechanism for the emergence of technological complexity. For a comprehensive understanding of the dynamics of technological complexity, it is important to take into account both the transversal and the longitudinal dimension of cumulativeness. Our study shows that cumulativeness increases along both these dimensions (for the considered technologies), which suggests an overall increase of technological complexity as well, yet this partially depends on the chosen measure of complexity. 
    \end{enumerate} 
    These insights lead to a number of implications for innovation policies which benefit from a detailed understanding of the cumulativeness of technologies, such as smart specialization. In their consideration of various technologies, these policies are advised to choose a comprehensive approach, including both the absolute cumulativeness as well the cumulativeness relative to the size of the knowledge base. Where the first is indicative for the overall difficulty of entry in a technology, the second is indicative for the relative difficulty of entry as compared to technologies with similar sized knowledge base. Furthermore, given that near future inventive activity (and with that knowledge output) allows for some estimation or planning, these policies are advised to additionally take into account the expected development of the cumulativeness of these technologies. Although these developments are sometimes considered a black box, we have demonstrated that the cumulativeness in fact develops rather predictably with the size of the knowledge base. In the longer run, policymakers should be aware that the rate of invention over time of a technology, usually a direct or indirect subject of policy interventions, is inversely related to the cumulativeness rates. Although the possible causality in this relation is as of yet unclear, the consequences are either way considerable. In the most extreme cases, it either implies a certain 'counter effect': that a substantial acceleration of inventive activities indirectly slows down the cumulativeness rate of a technology, or it implies that, despite efforts of acceleration or deceleration, the inventive rate is largely conditioned by the cumulativeness rate alone. 
    
\section{Acknowledgments}
We would like to thank Anton Pichler, Thomas Schaper and three anonymous reviewers for helpful comments on the script. The icons in Figures \ref{icon_US} and \ref{icon_EP} are made by Freepik, Eucalyp, fjstudio, Those Icons, Pixel perfect, Kiranshastry, Becris, Smashicons, Prosymbols and Good Ware from www.flaticon.com. This work was supported by NWO (Dutch Research Council) grant nr. 452-13-010
    
\printbibliography

\appendix

\section{Overview of the statistics}\label{appendix_3}

In this appendix we present in the statistical details the linear fits in Figure \ref{id_ipl_patents}, \ref{max_speed} and the \ref{PL_IKD}. 

\subsection{Linear fits id and ipl}
We fit the straight lines in Figure \ref{id_ipl_patents} for each technology by an OLS regression, the results are presented in Tables \ref{reg1},\ref{reg2},\ref{reg3} and \ref{reg4}. 
\begin{table}[!htbp] \centering 
  \caption{\textbf{Estimated linear models for Internal Dependence (US patents)}} 
  \label{reg1} 
  \resizebox{\textwidth}{!}{%
\begin{tabular}{@{\extracolsep{5pt}}lcccc} 
\\[-1.8ex]\hline 
\hline \\[-1.8ex] 
 & \multicolumn{4}{c}{\textit{Dependent variable:}} \\ 
\cline{2-5} 
\\[-1.8ex] & Nuclear Fission & Photovoltaics & Wind Turbines & Combustion Engines \\ 
\\[-1.8ex] & (1) & (2) & (3) & (4)\\ 
\hline \\[-1.8ex] 
 Patents & 0.0006$^{***}$ & 0.0005$^{***}$ & 0.0014$^{***}$ & 0.0002$^{***}$ \\ 
  & (0.000002) & (0.000001) & (0.000004) & (0.000000) \\ 
  & & & & \\ 
 Constant & 0.6462$^{***}$ & 1.4492$^{***}$ & 2.4157$^{***}$ & 0.2560$^{***}$ \\ 
  & (0.0032) & (0.0045) & (0.0096) & (0.0012) \\ 
  & & & & \\ 
\hline \\[-1.8ex] 
Observations & 3,595 & 9,066 & 3,979 & 6,068 \\ 
R$^{2}$ & 0.98 & 0.97 & 0.96 & 0.98 \\ 
Adjusted R$^{2}$ & 0.98 & 0.97 & 0.96 & 0.99 \\ 
Residual Std. Error & 0.096 (df = 3593) & 0.214 (df = 9064) & 0.303 (df = 3977) & 0.047 (df = 6066) \\ 
F Statistic & 166,232$^{***}$ (df = 1; 3593) & 340,069$^{***}$ (df = 1; 9064) & 108,235$^{***}$ (df = 1; 3977) & 422,204$^{***}$ (df = 1; 6066) \\ 
\hline 
\hline \\[-1.8ex] 
\textit{Note:}  & \multicolumn{4}{r}{$^{*}$p$<$0.1; $^{**}$p$<$0.05; $^{***}$p$<$0.01} \\ 
\end{tabular} }%
\end{table} 

\begin{table}[!htbp] \centering 
  \caption{\textbf{Estimated linear models for Internal Path Length (US patents)}} 
  \label{reg2} 
  \resizebox{\textwidth}{!}{%
\begin{tabular}{@{\extracolsep{5pt}}lcccc} 
\\[-1.8ex]\hline 
\hline \\[-1.8ex] 
 & \multicolumn{4}{c}{\textit{Dependent variable:}} \\ 
\cline{2-5} 
\\[-1.8ex] & Nuclear Fission & Photovoltaics & Wind Turbines & Combustion Engines \\ 
\\[-1.8ex] & (1) & (2) & (3) & (4)\\ 
\hline \\[-1.8ex] 
 Patents & 0.0029$^{***}$ & 0.0024$^{***}$ & 0.0044$^{***}$ & 0.0011$^{***}$ \\ 
  & (0.00001) & (0.000003) & (0.00001) & (0.000004) \\ 
  & & & & \\ 
 Constant & $-$0.4162$^{***}$ & 2.6356$^{***}$ & 0.9349$^{***}$ & $-$0.0644$^{***}$ \\ 
  & (0.0198) & (0.0139) & (0.0302) & (0.0127) \\ 
  & & & & \\ 
\hline \\[-1.8ex] 
Observations & 3,595 & 9,066 & 3,979 & 6,068 \\ 
R$^{2}$ & 0.96 & 0.99 & 0.97 & 0.94 \\ 
Adjusted R$^{2}$ & 0.96 & 0.99 & 0.97 & 0.94 \\ 
Residual Std. Error & 0.595 (df = 3593) & 0.664 (df = 9064) & 0.951 (df = 3977) & 0.496 (df = 6066) \\ 
F Statistic & 90,488$^{***}$ (df = 1; 3593) & 828,425$^{***}$ (df = 1; 9064) & 113,419$^{***}$ (df = 1; 3977) & 98,777$^{***}$ (df = 1; 6066) \\ 
\hline 
\hline \\[-1.8ex] 
\textit{Note:}  & \multicolumn{4}{r}{$^{*}$p$<$0.1; $^{**}$p$<$0.05; $^{***}$p$<$0.01} \\ 
\end{tabular} }%
\end{table} 

\begin{table}[!htbp] \centering 
  \caption{\textbf{Estimated linear models for Internal Dependence (European patents)}} 
  \label{reg3} 
  \resizebox{\textwidth}{!}{%
\begin{tabular}{@{\extracolsep{5pt}}lcccc} 
\\[-1.8ex]\hline 
\hline \\[-1.8ex] 
 & \multicolumn{4}{c}{\textit{Dependent variable:}} \\ 
\cline{2-5} 
\\[-1.8ex] & Id Nuclear Fission & Id Photovoltaics & Id Wind Turbines & Id Combustion Engines \\ 
\\[-1.8ex] & (1) & (2) & (3) & (4)\\ 
\hline \\[-1.8ex] 
 patents & 0.0011$^{***}$ & 0.0002$^{***}$ & 0.0008$^{***}$ & 0.0002$^{***}$ \\ 
  & (0.00001) & (0.000002) & (0.000003) & (0.000001) \\ 
  & & & & \\ 
 Constant & 0.0665$^{***}$ & 0.2522$^{***}$ & 0.1812$^{***}$ & 0.0690$^{***}$ \\ 
  & (0.0042) & (0.0029) & (0.0028) & (0.0009) \\ 
  & & & & \\ 
\hline \\[-1.8ex] 
Observations & 744 & 2,598 & 1,766 & 2,092 \\ 
R$^{2}$ & 0.95 & 0.81 & 0.98 & 0.97 \\ 
Adjusted R$^{2}$ & 0.95 & 0.81 & 0.98 & 0.9616 \\ 
Residual Std. Error & 0.057 (df = 742) & 0.073 (df = 2596) & 0.059 (df = 1764) & 0.021 (df = 2090) \\ 
F Statistic & 13,048$^{***}$ (df = 1; 742) & 11$^{***}$ (df = 1; 2596) & 76$^{***}$ (df = 1; 1764) & 52,396$^{***}$ (df = 1; 2090) \\ 
\hline 
\hline \\[-1.8ex] 
\textit{Note:}  & \multicolumn{4}{r}{$^{*}$p$<$0.1; $^{**}$p$<$0.05; $^{***}$p$<$0.01} \\ 
\end{tabular} }%
\end{table} 

\begin{table}[!htbp] \centering 
  \caption{\textbf{Estimated linear models for Internal Path Length (European patents)}} 
  \label{reg4} 
   \resizebox{\textwidth}{!}{%
\begin{tabular}{@{\extracolsep{5pt}}lcccc} 
\\[-1.8ex]\hline 
\hline \\[-1.8ex] 
 & \multicolumn{4}{c}{\textit{Dependent variable:}} \\ 
\cline{2-5} 
\\[-1.8ex] & Ipl Nuclear Fission & Ipl Photovoltaics & Ipl Wind Turbines & Ipl Combustion Engines \\ 
\\[-1.8ex] & (1) & (2) & (3) & (4)\\ 
\hline \\[-1.8ex] 
 patents & 0.0023$^{***}$ & 0.0008$^{***}$ & 0.0021$^{***}$ & 0.0004$^{***}$ \\ 
  & (0.00001) & (0.000003) & (0.00001) & (0.000002) \\ 
  & & & & \\ 
 Constant & $-$0.0473$^{***}$ & 0.2337$^{***}$ & 0.0913$^{***}$ & 0.0274$^{***}$ \\ 
  & (0.0046) & (0.0046) & (0.0070) & (0.0023) \\ 
  & & & & \\ 
\hline \\[-1.8ex] 
Observations & 744 & 2,598 & 1,766 & 2,088 \\ 
R$^{2}$ & 0.98 & 0.96 & 0.98 & 0.95 \\ 
Adjusted R$^{2}$ & 0.98 & 0.96 & 0.98 & 0.95 \\ 
Residual Std. Error & 0.062 (df = 742) & 0.117 (df = 2596) & 0.147 (df = 1764) & 0.052 (df = 2086) \\ 
F Statistic & 47,153$^{***}$ (df = 1; 742) & 69$^{***}$ (df = 1; 2596) & 92$^{***}$ (df = 1; 1764) & 37,728$^{***}$ (df = 1; 2086) \\ 
\hline 
\hline \\[-1.8ex] 
\textit{Note:}  & \multicolumn{4}{r}{$^{*}$p$<$0.1; $^{**}$p$<$0.05; $^{***}$p$<$0.01} \\ 
\end{tabular} }%
\end{table} 

\subsection{Maximum Internal Path Length}
In Table \ref{regmax} we present the results of the linear regressions in Figure \ref{max_speed}. 
\begin{table}[!htbp] \centering 
 \caption{\textbf{Development of Maximum Internal Path Length (mipl) (US patents)}} 
  \label{regmax} 
   \resizebox{\textwidth}{!}{%
\begin{tabular}{@{\extracolsep{5pt}}lcccc} 
\\[-1.8ex]\hline 
\hline \\[-1.8ex] 
 & \multicolumn{4}{c}{\textit{Dependent variable:}} \\ 
\cline{2-5} 
\\[-1.8ex] & Mipl Nuclear Fission & Mipl Photovoltaics & Mipl Wind Turbines & Mipl Combustion Engines \\ 
\\[-1.8ex] & (1) & (2) & (3) & (4)\\ 
\hline \\[-1.8ex] 
patents & 0.0062$^{***}$ & 0.0046$^{***}$ & 0.0089$^{***}$ & 0.0021$^{***}$ \\ 
  & (0.0002) & (0.0001) & (0.0003) & (0.0002) \\ 
  & & & & \\ 
 Constant & 1.54$^{***}$ & 8.31$^{***}$ & 2.07$^{***}$ & 3.34 \\ 
  & (2.6496) & (0.4073) & (1.6830) & (0.7232) \\ 
  & & & & \\ 
\hline \\[-1.8ex] 
Observations & 28 & 49 & 33 & 17 \\ 
R$^{2}$ & 0.97 & 0.98 & 0.97 & 0.91 \\ 
Adjusted R$^{2}$ & 0.97 & 0.98 & 0.97 & 0.91 \\ 
Residual Std. Error & 1.48 (df = 26) & 1.84 (df = 47) & 1.65 (df = 31) & 1.54 (df = 15) \\ 
F Statistic & 814$^{***}$ (df = 1; 26) & 2,836$^{***}$ (df = 1; 47) & 1,062$^{***}$ (df = 1; 31) & 157$^{***}$ (df = 1; 15) \\ 
\hline 
\hline \\[-1.8ex] 
\textit{Note:}  & \multicolumn{4}{r}{$^{*}$p$<$0.1; $^{**}$p$<$0.05; $^{***}$p$<$0.01} \\ 
\end{tabular}  }%
\end{table} 

\subsection{Cross technology comparisons with additional technologies}
We present in Table \ref{cross} the outcomes of the pairwise regressions in Figure \ref{PL_IKD}. Earlier we included more detailed information on the four focus technologies in Table \ref{desstat}. A more detailed description of the other 20 technologies in Table \ref{PL_IKD}, including their CPC classification and number of granted patents with earliest filing date before the year 2009 can be found in Table \ref{addtech}. While the choice for these particular technologies was mostly arbitrary, we took care to include technologies from each main CPC section (indicated by first letter A,B,C,D,E,F,Y) and from mostly different CPC subclasses (indicated by first 4 symbols e.g. C10J). To further limit the scope of the technologies we selected the technologies on the CPC groups and subgroup level. Even though this selection of the group and subgroups was mostly arbitrary, we took into account that we require a substantial number of patents for each technology (>200). 

\begin{table}[!htbp] \centering 
  \caption{\textbf{Pair wise regression between id and ipl for EP and US patents}} 
  \label{cross} 
\begin{tabular}{@{\extracolsep{5pt}}lcccc} 
\\[-1.8ex]\hline 
\hline \\[-1.8ex] 
 & \multicolumn{4}{c}{\textit{Dependent variable:}} \\ 
\cline{2-5} 
\\[-1.8ex] & \multicolumn{2}{c}{id US patents} & \multicolumn{2}{c}{ipl EP patents} \\ 
\\[-1.8ex] & (1) & (2) & (3) & (4)\\ 
\hline \\[-1.8ex] 
 id EP patents & 3.284$^{***}$ &  & 4.625$^{***}$ &  \\ 
  & (0.870) &  & (1.449) &  \\ 
  & & & & \\ 
 ipl US patents &  & 0.127$^{***}$ &  & 0.264$^{***}$ \\ 
  &  & (0.043) &  & (0.057) \\ 
  & & & & \\ 
 Constant & 1.509$^{*}$ & 2.278$^{***}$ & $-$0.256 & $-$0.599 \\ 
  & (0.828) & (0.795) & (1.378) & (1.045) \\ 
  & & & & \\ 
\hline \\[-1.8ex] 
Observations & 24 & 24 & 24 & 24 \\ 
R$^{2}$ & 0.39 & 0.28 & 0.32 & 0.50 \\ 
Adjusted R$^{2}$ & 0.37 & 0.25 & 0.29 & 0.47 \\ 
Residual Std. Error (df = 22) & 1.49 & 1.62 & 2.48 & 2.13 \\ 
F Statistic (df = 1; 22) & 14.25$^{***}$ & 8.73$^{***}$ & 10.19$^{***}$ & 21.76$^{***}$ \\ 
\hline 
\hline \\[-1.8ex] 
\textit{Note:}  & \multicolumn{4}{r}{$^{*}$p$<$0.1; $^{**}$p$<$0.05; $^{***}$p$<$0.01} \\ 
\end{tabular} 
\end{table} 
\begin{table}[!htbp] \centering 
\caption{\textbf{Invention rate over time for the id coefficients $q$} The log of US invention rate (patent/year) is regressed for the log of the id coefficient $q$ (reference/patent$^2$) for the 24 technologies in Tables \ref{desstat} and \ref{addtech}, for both the US and EP patents. The $q$ coefficients are calculated by dividing the total references by the total patents squared.} 
  \label{rate_regress} 
\begin{tabular}{@{\extracolsep{5pt}}lcc} 
\\[-1.8ex]\hline 
\hline \\[-1.8ex] 
 & \multicolumn{2}{c}{\textit{Dependent variable:}} \\ 
\cline{2-3} 
\\[-1.8ex] & Log Invention Rate US & Log Invention rate EP \\ 
\\[-1.8ex] & (1) & (2)\\ 
\hline \\[-1.8ex] 
 Log US id coefficient q  & $-$0.595$^{***}$ &  \\ 
  & (0.166) &  \\ 
  & & \\ 
 Log EP id coefficient q &  & $-$0.867$^{***}$ \\ 
  &  & (0.186) \\ 
  & & \\ 
 Constant & 0.302 & $-$2.463$^{*}$ \\ 
  & (1.174) & (1.398) \\ 
  & & \\ 
\hline \\[-1.8ex] 
Observations & 24 & 24 \\ 
R$^{2}$ & 0.368 & 0.496 \\ 
Adjusted R$^{2}$ & 0.339 & 0.473 \\ 
Residual Std. Error (df = 22) & 0.625 & 0.579 \\ 
F Statistic (df = 1; 22) & 12.791$^{***}$ & 21.635$^{***}$ \\ 
\hline 
\hline \\[-1.8ex] 
\textit{Note:}  & \multicolumn{2}{r}{$^{*}$p$<$0.1; $^{**}$p$<$0.05; $^{***}$p$<$0.01} \\ 
\end{tabular} 
\end{table}

\begin{table}[!htbp]
\caption{\textbf{CPC code and description of additional technologies}}
\label{addtech}
\resizebox{\textwidth}{!}{%
\begin{tabular}{lllll}
\hline
\textbf{Technology name} &
  \textbf{CPC description} &
  \textbf{CPC code} &
  \textbf{\begin{tabular}[c]{@{}l@{}}\# US granted \\ patents filed \\ earliest\textless{}2009\end{tabular}} &
  \textbf{\begin{tabular}[c]{@{}l@{}}\# EP granted \\ patents filed \\ earliest\textless{}2009\end{tabular}} \\ \hline
  
  animal housings &
  \begin{tabular}[c]{@{}l@{}}animal husbandry: \\ equipment for housing animals\end{tabular} &
  A01K  1 &
  5974 &
  851 \\ \hline
soil reclamation &
  reclamation of contaminated soil &
  B09C  1 &
  2118 &
  608 \\ \hline
fusion polypeptide &
  fusion polypeptide &
  C07K2319 &
  7847 &
  4049 \\ \hline  
film sheet manufacturing &
  \begin{tabular}[c]{@{}l@{}}manufacture of articles or shaped materials \\ containing macromolecular  substances: \\ films or sheets\end{tabular} &
  C08J   5/18 &
  3570 &
  2114 \\ \hline 
filament manufacturing &
  \begin{tabular}[c]{@{}l@{}}general methods for the manufacture of \\ artificial filaments or the like\end{tabular} &
  D01F    1 &
  1383 &
  570 \\ \hline
  laundry dryers &
  \begin{tabular}[c]{@{}l@{}}domestic laundry dryers\end{tabular} &
  D06F  58 &
  2701 &
  839 \\ \hline
  homopolymers & \begin{tabular}[c]{@{}l@{}} Homopolymers and copolymers of \\ unsaturated aliphatic hydrocarbons having \\ only one carbon-tocarbon
double bond\end{tabular} & C08F 10 & 8554 & 3461 \\ \hline
soil-shifting machines &
  soil-shifting machines &
  E02F   9 &
  6456 &
  1560 \\ \hline
exhaust purifiers &
  \begin{tabular}[c]{@{}l@{}}exhaust or silencing apparatus having means \\ for purifying, rendering  innocuous, or otherwise \\ treating exhaust for extinguishing sparks\end{tabular} &
  F01N   3/2066 &
  766 &
  629 \\ \hline
combustibles supply control &
  \begin{tabular}[c]{@{}l@{}}electrical control of supply of \\ combustible mixture or its constituents\end{tabular} &
  F02D   41 &
  18187 &
  6816 \\ \hline
  gearing control functions & \begin{tabular}[c]{@{}l@{}}Control functions within control \\units of changespeed- or reversing-gearings \\ for conveying rotary motion\end{tabular} & F16H  61 & 15253 & 4463 \\ \hline
sensing record carriers &
  \begin{tabular}[c]{@{}l@{}}methods or arrangements for sensing\\ record carriers,\end{tabular} &
  G06K  7 &
  11142 &
  3316 \\ \hline
object actuated machines &
  \begin{tabular}[c]{@{}l@{}}mechanisms actuated by objects other \\ than coins to free or to actuate vending, \\ hiring, coin or paper currency dispensing \\ or refunding apparatus\end{tabular} &
  G07F   7 &
  6799 &
  2214 \\ \hline
control signal transmission &
  \begin{tabular}[c]{@{}l@{}}arrangements for transmitting signals \\ characterised by the use of a\\ wireless electrical link\end{tabular} &
  G08C   17/02 &
  907 &
  313 \\ \hline
speech recognition procedures &
  \begin{tabular}[c]{@{}l@{}}procedures used during a speech recognition \\ process, e.g. man-machine  dialogue\end{tabular} &
  G10L   15/22 &
  919 &
  223 \\ \hline
information storage editing &
  \begin{tabular}[c]{@{}l@{}}editing; indexing; addressing; timing or \\ synchronising; monitoring;  measuring \\ tape travel\end{tabular} &
  G11B   27 &
  14607 &
  3501 \\ \hline
lithium battery manufacturing &
  \begin{tabular}[c]{@{}l@{}}manufacturing of secondary cells, \\ accumulators with non-aqueous electrolyte, \\ lithium accumulators\end{tabular} &
  H01M  10/052 &
  5550 &
  2203 \\ \hline
predictive video signal coding &
  \begin{tabular}[c]{@{}l@{}}methods or arrangements for coding, decoding, \\ compressing or decompressing digital video \\ signals using transform coding in combination \\ with predictive coding\end{tabular} &
  H04N   19/61 &
  5411 &
  1371 \\ \hline
electroluminescent light sources &
  electroluminescent light sources &
  H05B   33 &
  6933 &
  1848 \\ \hline
non-fossil fuels &
  \begin{tabular}[c]{@{}l@{}}technologies for the production of fuel \\ of nonfossil origin\end{tabular} &
  Y02E  50 &
  3627 &
  1549 \\ \hline
\end{tabular}%
}
\end{table}

Finally we include in Table \ref{super_class} the grouping of CPC classes into a more aggregated level of technological fields. We also include the number of unique patents in these fields. 
\begin{table}[]
\centering
\caption{Aggregration of CPC classes and number of unique patents}
\label{super_class}
\resizebox{\textwidth}{!}{%
\begin{tabular}{|l|l|l|l|}
\hline
\textbf{AGGREGATED CLASS} & \textbf{CPC classes} & \textbf{US patents  earliest filed <2009} & \textbf{EP patents earliest filed <2009} \\ \hline
AGRICULTURE                     & A0              & 199829 & 62274   \\ \hline
FOODSTUFFS \& TOBACCO            & A21-A24         & 99283  & 99283   \\ \hline
WEARABLES                       & A41-A46         & 158168 & 31274   \\ \hline
FURNITURE \& DOMESTIC ARTICLES & A47             & 198137 & 35881   \\ \hline
HEALTH \& WELLBEING              & A61-A63         & 551343 & 1057394 \\ \hline
SEPARATING \& MIXING PROCESSES   & B01-B09         & 299444 & 188607  \\ \hline
SHAPING OF MATERIALS            & B21-B33         & 569035 & 257470  \\ \hline
PRINTING \& DECORATION           & B41-B44         & 164956 & 106699  \\ \hline
RAILROADS \& SHIPS           & B61,B63         & 113111 & 9208  \\ \hline 
AVIATION                        & B64         & 50379 & 7350  \\ \hline 
OTHER, VEHICLES RELATED          & B60,B62         & 364114 & 76396  \\ \hline
PACKING \& TRANSPORTING          & B65-B68         & 343185 & 124825  \\ \hline
MICRO \& NANOTECHNOLOGY          & B81,B82         & 33116  & 14804   \\ \hline
INORGANIC CHEMISTRY             & C01             & 64289  & 29767   \\ \hline
OTHER CHEMICAL INDUSTRIES       & C02-C06,C13,C14 & 130118 & 69540   \\ \hline
ORGANIC CHEMISTRY               & C07         & 335300
 & 105630  \\ \hline
MACROMOLECULES               &  C08         & 217371 & 72606  \\ \hline
PAINTS                          & C09             & 113624 & 124642  \\ \hline
PETROLEUM, GAS \& COKE           & C10             & 80664  & 31815   \\ \hline
BIOCHEMISTRY                    & C11,C12         & 110121 & 190851  \\ \hline
METALLURGY                      & C21-C25         & 143398 & 71321   \\ \hline
OTHER CHEMISTRY                 & C30,C40         & 14622  & 9095    \\ \hline
TEXTILES                        & D01-D10         & 174651 & 63344   \\ \hline
PAPER-MAKING                    & D21             & 35187  & 29176   \\ \hline
BUILDING                        & E01-E06         & 292294 & 64659   \\ \hline
MINING                          & E21             & 81643  & 20124   \\ \hline
ENGINES \& PUMPS                 & F01-F05         & 276811 & 151200  \\ \hline
MECHANICAL ENGINEERING          & F15-F17         & 392610 & 145528  \\ \hline
LIGHTING                        & F21             & 40427  & 8417    \\ \hline
HEATING                         & F22-F28         & 229621 & 63076   \\ \hline
WEAPONS                         & F41,F42         & 61036  & 11039   \\ \hline
MEASURING \& TESTING             & G01,G04         & 452562 & 229884  \\ \hline
OPTICS \& PHOTOGRAPHY            & G02,G03         & 314480 & 159094  \\ \hline
COMPUTATION \& CONTROLLING       & G05-G07         & 493428 & 200383  \\ \hline
SOUND, SIGNALLING \& INFORMATION & G08-G16         & 331302 & 133285  \\ \hline
NUCLEONICS                      & G21             & 21952  & 9357    \\ \hline
ELECTRIC ELEMENTS \& CIRCUITRY   & H01-H03, H05    & 846273 & 351941  \\ \hline
ELECTRIC POWER GENERATION       & H02             & 148028 & 52661   \\ \hline
TELECOMMUNICATION               & H04             & 456994 & 429038  \\ \hline
CLIMATE CHANGE MITIGATION       & Y02             & 229074 & 131096  \\ \hline
\end{tabular}%
}
\end{table}
\section{Evaluating the distribution fits} \label{appendix_4}
\begin{figure}
\centering
\caption{\textbf{Probability plots for the backward link distributions (US patents)} We plot the probability plots for the backward link distributions for the four focus technologies, for the geometric distribution $G(\rho)$, the normal distribution $N(\mu,\sigma^2)$ and the binomial distribution $B(n,p)$. The parameters $\rho, \mu, \sigma^2$ and $p$ are optimized to obtain the best fit.}
\includegraphics[width=\textwidth]{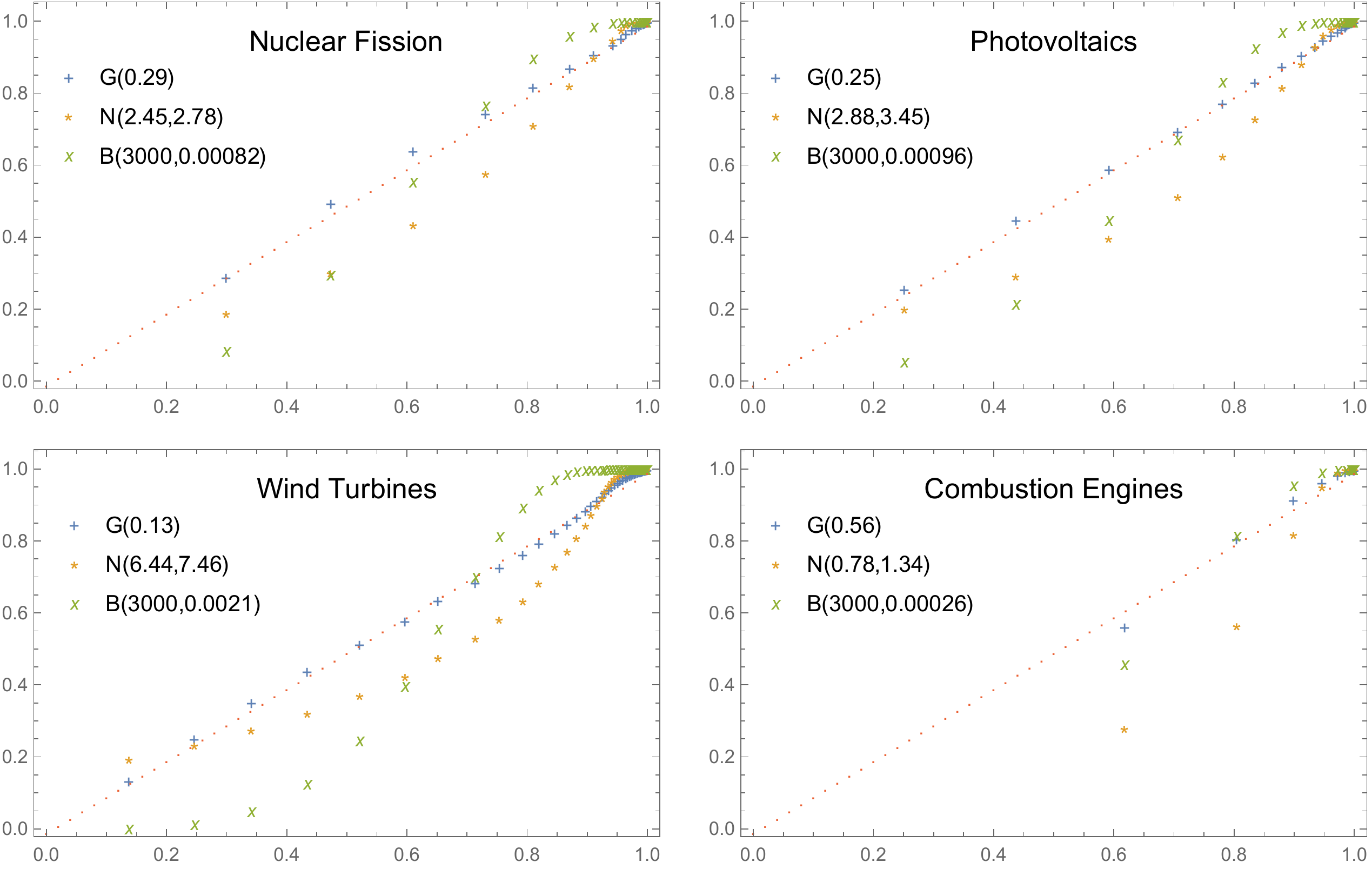}
\label{id_prob_plot}
\end{figure}

In this appendix we discuss the fits of the distributions in Figures \ref{id_dis} and \ref{ipl_dis}. For the distributions where less data is available (i.e. n=1000, 2000), $\chi^2$ tests indicate there is not enough evidence to reject the null-hypothesis that the backward link distributions are described by geometric distributions with parameters from Table \ref{coefficients}. However, for larger $n$, the $p$-values quickly get very small for virtually any distribution we try, which suggests that the $\chi^2$ test is rather strict for our purpose. Instead we therefore consider probability plots instead, where we compare the performance of the predicted distribution to a number of other possible candidates, such as the binomial distribution for the backward links and the normal distribution for the path length. The x-value of each point in the probability plot represent the empirical probability of a certain occurrence and its y-value represent the predicted probability of its occurrence. The closer the points to the $x=y$ line, the better the distribution fit therefore. The Figure \ref{id_prob_plot} shows the probability plots for the (empirical) backward link distribution of the four focus technologies and three candidate distributions: the geometric, normal and binomial distributions. We choose to show the $n=3000$ case, yet the other cases are largely comparable. The parameters of each distribution are chosen such that the fit with the empirical distribution is optimized. We observe that the geometric distribution is for all technologies very close to the $x=y$ line, more so than the other distributions. 

\begin{figure}
\centering
\caption{\textbf{Probability Plots for the path length distributions (US patents)} We plot the probability plots for the path length distributions for the four focus technologies, for the Poisson distribution $P(\eta)$, the normal distribution $N(\mu,\sigma^2)$ and the binomial type of distribution $B(n')$ from Equation \ref{main7}, (where the values of $n'$ correspond to those in in Figure \ref{max_speed}). The parameters $\eta, \mu, \sigma^2$ and are optimized to obtain the best fit. Each distributions is plotted for $n=3000$, except for wind turbines $n=2000$.}
\includegraphics[width=\textwidth]{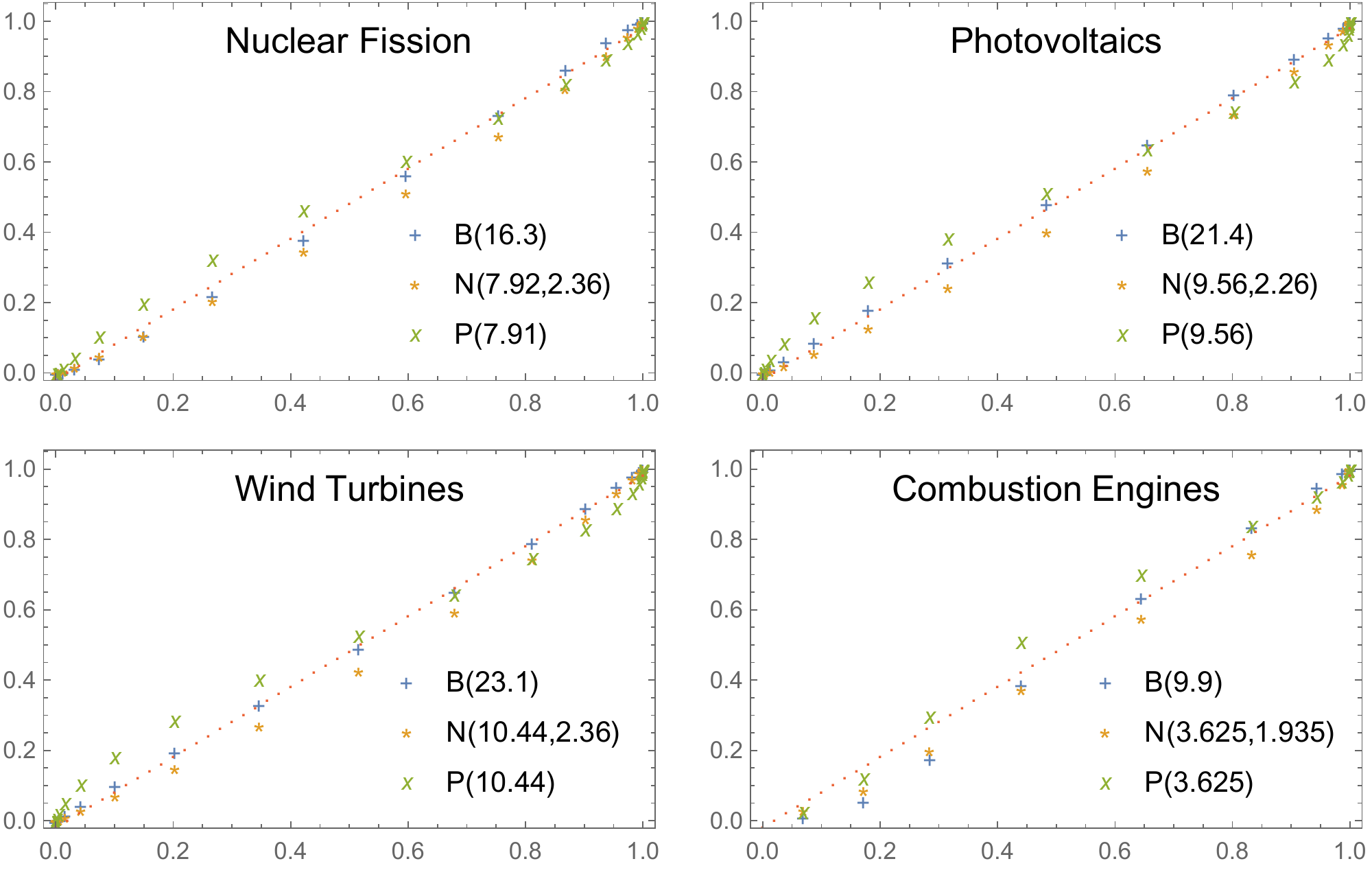}
\label{ipl_prob_plot}
\end{figure}

Similarly, Figure \ref{ipl_prob_plot} shows the probability plots for the path length distribution and three candidate distributions: the Poisson, normal and binomial type of distribution of Equation \ref{main7}\footnote{We choose again the $n=3000$ cases, except for computational reasons we chose $n=2000$ for wind turbines.}. For the path length distributions too we observe that the distribution from Equation \ref{main7} is generally close to the $x=y$ line for each technology. Only for the lower path length values of combustion engines this distribution deviates slightly more than the other distributions, yet overall it presents still the best fit. 
Note that the quality of fits provided by the geometric distributions in Figure \ref{id_prob_plot} and the distribution from Equation \ref{main7} in Figure \ref{ipl_prob_plot} (both single parameter distributions) is quite remarkable, especially in comparison to the normal distribution, which allows us to fit two parameters instead. 

\section{Examiner versus applicant citations}\label{appendix_5}  

In most patent offices, a citation can be introduced both by the applicant and the patent examiner. The citations added by the applicant are often perceived to be as the better indicator for knowledge flows \parencite{criscuolo_does_2008,jaffe_nbersloan_2000}, yet the difference between between the types of citation is/was not always recorded for each patents office, not the least being USPTO before the year 2000. This research therefore uses a general citation instead to represent a knowledge flow. As a justification for this choice we investigate in this appendix (when possible) the similarity between the knowledge dynamics based on applicant citations ("type APP") to the overall dynamics.  

\begin{figure}
\centering
\begin{minipage}{.48\linewidth}
  \includegraphics[width=\linewidth]{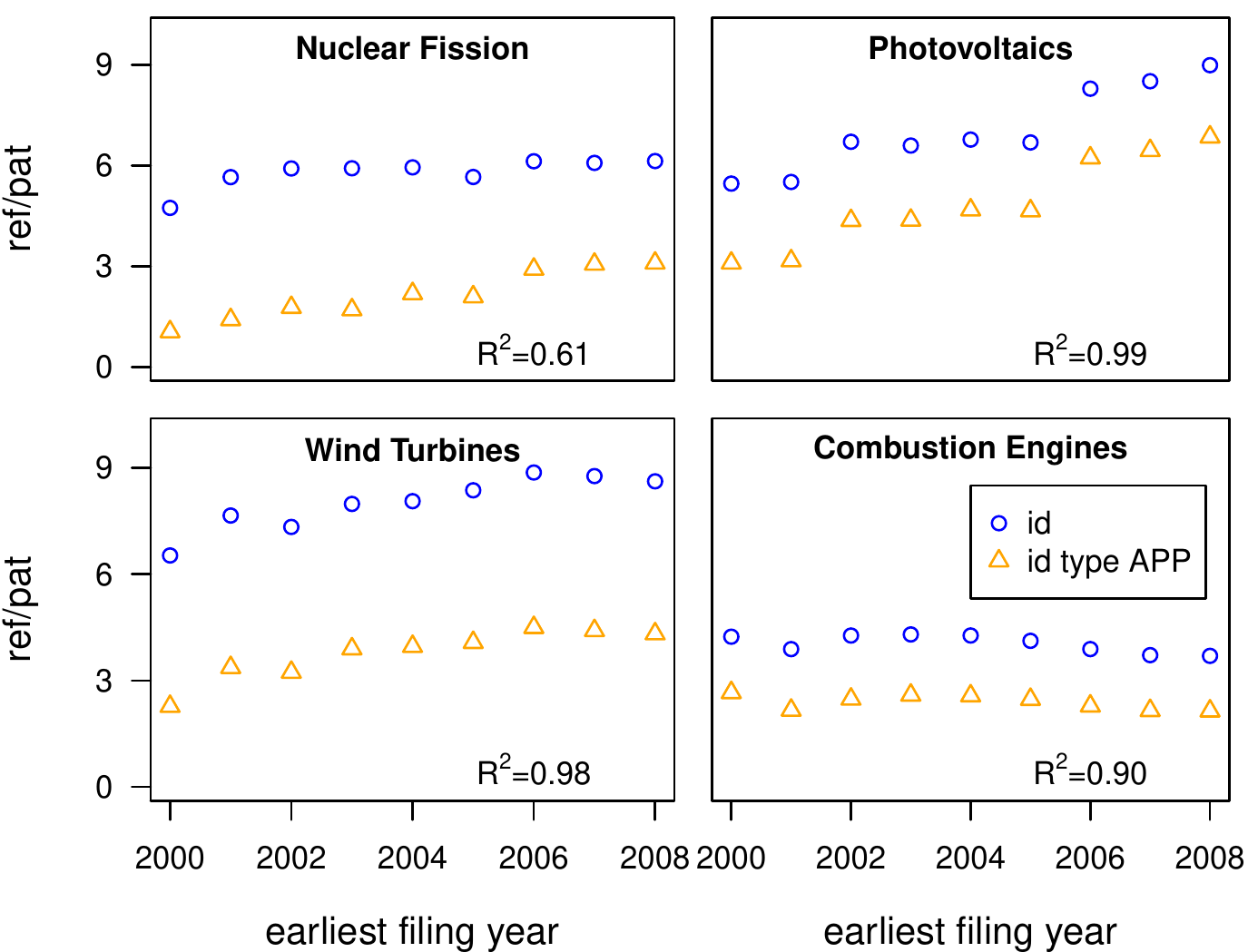}
  \caption{\textbf{Type APP id for US patents} We plot the 'normal' id and the id based on type applicant citations (type APP) for the earliest filing year. Both develop largely similar, which is also reflected by the relatively high correlation coefficients $R^2$}
\label{app_US}
\end{minipage}
\hspace{.02\linewidth}
\begin{minipage}{.48\linewidth}
  \includegraphics[width=\linewidth]{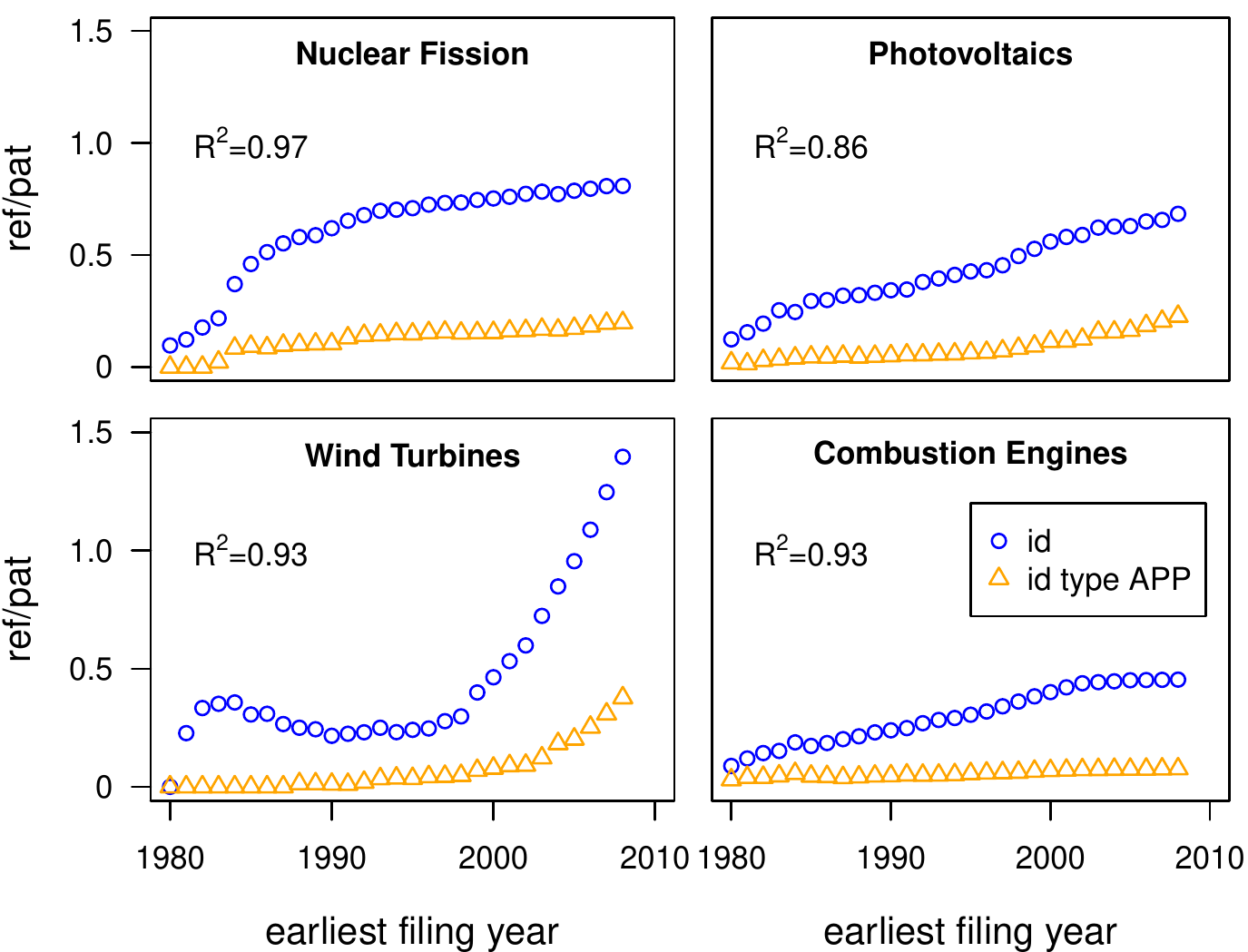}
  \caption{\textbf{Type APP id for EP patents} Similar to Figure \ref{app_US} but then for EP patents. Contrary to the US data, the distinction between applicant and examiner citations is recorded for the European data for the entire period of operation of the EPO.}
\label{app_EP}
\end{minipage}
\end{figure}

\begin{figure}
\centering
\caption{\textbf{Id versus id type APP for 24 technologies} In the left panel we compare the two id's for the US patents and on the right for European patents. Note that both axes are logarithmic, hence the fitted line is a power law.}
\includegraphics[width=\textwidth]{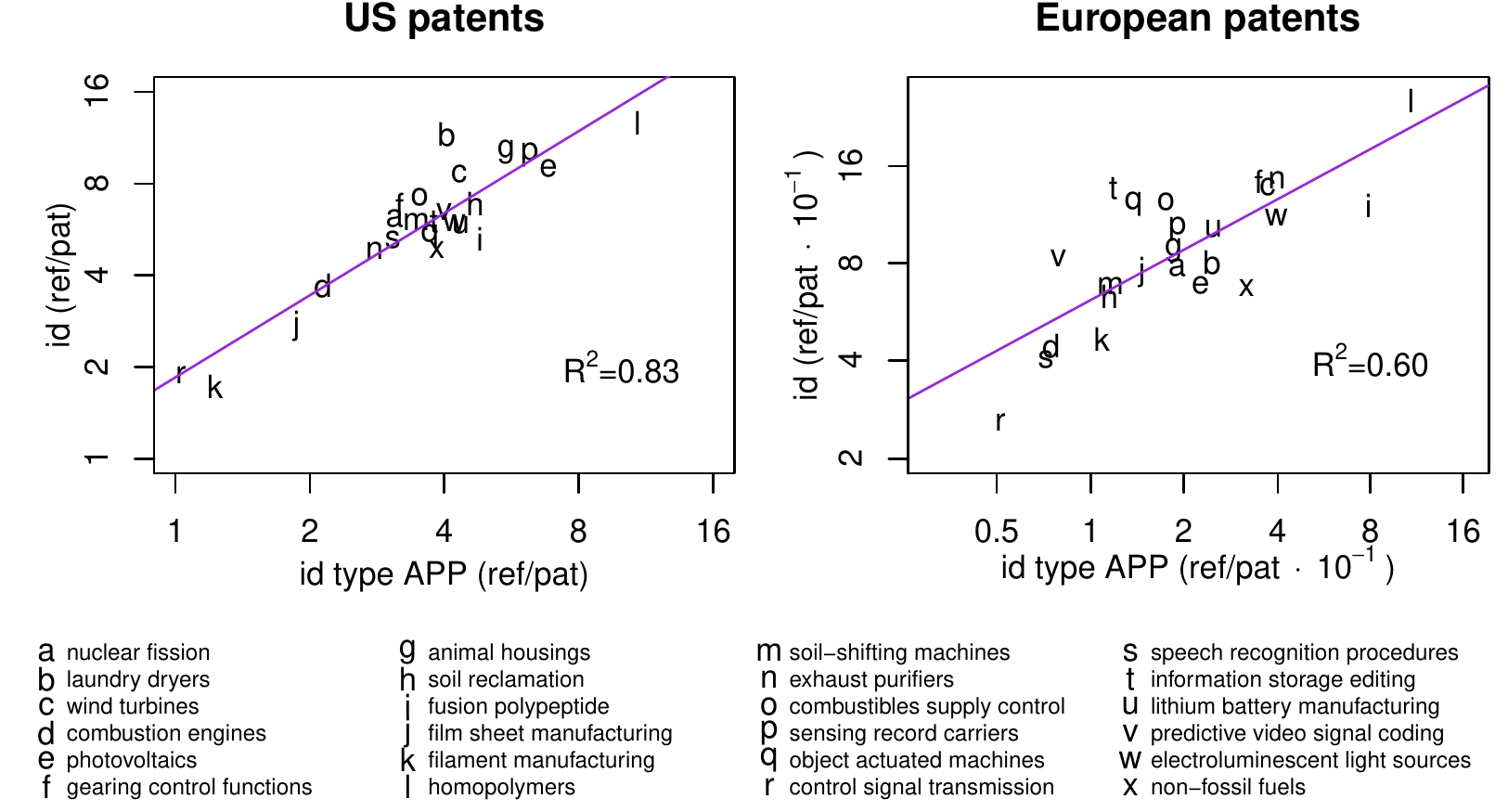}
\label{APP_cross}
\end{figure}

In line with \parencite{azagra-caro_examiner_2018, criscuolo_does_2008}, we determine for the European patents the type APP citations as those with Patstat's \textit{citn\_categ}='D'. For the US patents the type APP citations are selected as those with Patstat's \textit{citn\_origin}='APP'.\footnote{We supplemented the 2019 Patstat dataset with the 2018 Patstat dataset, which contains a more complete recording of the citn\_categ 'D'.} As mentioned earlier, for the US patents we only include data from about 2000 onward, as only after 2000 the distinction was made between examiner and applicant citations by USPTO \parencite{office_espacenet_nodate}. In Figures \ref{app_EP} and \ref{app_US} we plot the internal dependence based on Type APP citations both for the US and EP patents for different years. We observe that both dependences develop rather similarly over time, which is confirmed by the high correlation coefficients $R^2$ for each technology. The type APP id's are consistently a fraction lower as the citations added by the application are a subset of the total citations. Where for the EP patents the type APP citations are about a quarter of the total citations, for the US patents it is about a half. For both cases however the fraction varies somewhat per technology. We therefore systematically compare both id's for the entire set of 24 technologies in Figure \ref{APP_cross}. While a power law provides the best fit (as illustrated in \ref{APP_cross}), the relation is also rather well fitted by a simple linear relation. Regardless of the exact form, both id's are rather closely (and positively) related across technologies. As the id is closely related to the internal path length, this suggests that we can draw a similar conclusion for the latter indicator.    

In conclusion, while there is no guarantee that both types of citation result in similar knowledge dynamics, these results suggest that the knowledge dynamics based on type app citations are closely (and positively) related to the knowledge dynamics based on general citations. 

\section{Mathematical appendix}\label{math}

In this appendix we explain the number of mathematical derivations appear in Section \ref{ipl_section} in more detail. We start by explaining how the assumption that $n_{0}=rn$ is compatible with the geometric distribution found for the number of backward links in Section \ref{id_section} if $q$ is small compared to $m_{0}$. Using that the probability to obtain an initial node is $P_{n}(m=0)=\frac{1}{qn+m_{1}}$, we note that the expected number of initial nodes $\langle n_{0}\rangle$ after $n$ inventions is
\begin{align}
\sum_{n'=1}^{n}P_{n}(m=0)&=\frac{1}{q}H(n+\frac{m_{1}}{q})-\frac{1}{q}H(\frac{m_{1}}{q}) \\ &\approx \frac{1}{q}\log(n+\frac{m_{1}}{q})-\frac{1}{q}\log(\frac{m_{1}}{q}) \\
& \approx \frac{1}{q}\log(1+\frac{qn}{m_{1}}),
\end{align}
where we approximated the harmonic numbers $H(n)$ by logarithms. When $\frac{qn}{m_{1}}$ is small, i.e. when $q<<m_{1}$, we can approximate the last expression as $\langle n_{0}\rangle\approx \frac{n}{m_{1}}$. This suggests that, for $q<<m_{1}$  we can approximate the coefficient $r\approx \frac{1}{m_{1}}=\frac{1}{m_{0}+1}$. 

Next we discuss the steps leading to Equation \ref{main2},\ref{main4},\ref{slope} and \ref{main6}. To see that the expression in Equation \ref{main2} satisfies Equation \ref{main3}, first note that $\binom{n}{k}=0$ for $k>n$ and that $r\binom{n}{1}=rn$, as the initial conditions require. Then, start from the recursive property of the binomial coefficient $\binom{n+1}{k+1}-\binom{n}{k+1}=\binom{n}{k}$ and multiply left and right by $rq^{k}$. We then obtain
\begin{align}
    rq^{k}\binom{n+1}{k+1}-rq^{k}\binom{n}{k+1}&=rqq^{k-1}\binom{n}{k} \\
    f_{k}(n+1)-f_{k}(n)&=qf_{k-1}(n),
\end{align}
which is Equation \ref{main3}. To sum $f_{k}(n)$ from $k$ to $n$ (we need not sum further as all $f_{k}(n)=0$ for $k>n$), we can use the binomial theorem $\sum_{k=0}^{n}\binom{n}{k}x^{k}y^{n-k}=(x+y)^{n}$, which counts for any real (or complex) number $x$ and $y$. Taking $x=q$ and $y=1$ we get 
\begin{align}
    \sum_{k=0}^{n}\binom{n}{k}q^{k}&=(q+1)^{n} \\
    \sum_{k=1}^{n}\binom{n}{k}q^{k}&=(q+1)^{n}-1 \\
    r\sum_{k=0}^{n}\binom{n}{k+1}q^{k+1}&=r(q+1)^{n}-r \\ \label{eq17}
    \sum_{k=0}^{n}f_{k}(n)&=\frac{r(q+1)^{n}-r}{q}.
\end{align}
To obtain the expression in Equation \ref{main4} we divide Equation \ref{main2} by the right-hand side of Equation \ref{eq17}. To obtain Equation \ref{slope}, we have to calculate $\langle k \rangle=\sum_{k=0}^{n}k\tilde{f}_{k}(n)$. We first calculate  $\sum_{k=0}^{n}kf_{k}(n)=r\sum_{k=0}^{n}k\binom{n}{k+1}q^{k}$ instead. Note that this is exactly the expression we get if we differentiate  $r\sum_{k=0}^{n}\binom{n}{k+1}q^{k}$, i.e. the left-hand side of Equation \ref{eq17}, with respect to $q$ and then multiply once by $q$. Equivalently, we do this to the right-hand side of Equation \ref{eq17}, obtaining
\begin{equation}
    \sum_{k=0}^{n}kf_{k}(n)=q\frac{\partial}{\partial q}\frac{r(q+1)^{n}-r}{q}=r\bigg(n(1+q)^{n-1}-\frac{(1+q)^{n}-1}{q}\bigg).
\end{equation}
To obtain $\sum_{k=0}^{n}k\tilde{f}_{k}(n)$ from $\sum_{k=0}^{n}kf_{k}(n)$, we divide simply by the total number of paths (the right-hand side of Equation \ref{eq17}),  
\begin{align}
    \sum_{k=0}^{n}k\tilde{f}_{k}(n)&=\frac{qn(1+q)^{n-1}}{(1+q)^{n}-1}-1\\
    &=\frac{nq}{q+1}\cdot\frac{1}{1-(1+q)^{-n}}-1.
\end{align}
For large $n$, this expression quickly goes to $\frac{nq}{q+1}-1$. Hence we obtain Equation \ref{slope} for $k_{0}=-1$.

Finally, we note that solving the expression $f'_{k+1}(n'+1)-f'_{k+1}(n)=f'_{k}(n')$ goes completely analogous, (where $f\rightarrow f'$ and $n\rightarrow n'$), except that there is no coefficient $q$ (or the coefficient is '1'). Equivalently, therefore, we redo the approach in this appendix and do the substitutions $f\rightarrow f'$, $n\rightarrow n'$ and $q\rightarrow 1$. Substitutions in Equation \ref{main4} and \ref{slope} directly lead to respective Equations \ref{main7} and \ref{main6}. Note that the initial conditions change accordingly: (a) with a maximum speed $v=1/\delta n$, $f'_{k}(n')=0$ for $k>nv=n'$ and (b) at $n'$ nodes we expect to have $n'r$ initial nodes. 
\end{document}